\documentclass{aa}  

\usepackage{graphicx}
\usepackage{txfonts}
\usepackage{amsmath}

\usepackage[hyperfootnotes=true, colorlinks, citecolor=blue, linkcolor=blue, linktocpage]{hyperref}
\begin{document}

%\title{A universal approach to optimal frequency identification using nonparametric kernel regression}
\title{FINKER: Frequency Identification through Nonparametric KErnel Regression in astronomical time series}
%\title{FINKER: Frequency Identification through Nonparametric KErnel Regression}
%\title{PINCER:	Period Identification NonparametriC kErnel Regression}

%Cole's contribution:
%\title{NKR Periodogram: Identifying periodic signals with the Nonparametric Kernel Regression Periodogram}

\author{
    F. Stoppa\inst{1,2}\thanks{Equal contribution.}
    \and
    C. Johnston\inst{1,3,4}\protect\footnotemark[1]
    \and
    E. Cator\inst{2}
    \and
    G. Nelemans\inst{1,4,5}
    \and
    P.J. Groot\inst{1,6,7}
}

\institute{
    Department of Astrophysics/IMAPP, Radboud University, PO Box 9010,
    6500 GL Nijmegen, The Netherlands
    \and
    Department of Mathematics/IMAPP, Radboud University, PO Box 9010, 6500 GL Nijmegen, The Netherlands
    \and
    Max-Planck-Institut f\"ur Astrophysik, Karl-Schwarzschild-Stra{\ss}e 1, 85741 Garching  bei M\"unchen, Germany
    \and
    Institute of Astronomy, KU Leuven, Celestijnenlaan 200D, B-3001 Leuven, Belgium
    \and
    SRON, Netherlands Institute for Space Research, Sorbonnelaan 2, NL-3584 CA Utrecht, The Netherlands
    \and
    Department of Astronomy and Inter-University Institute for Data Intensive Astronomy, University of Cape Town, Private Bag X3, Rondebosch, 7701, South Africa
    \and
    South African Astronomical Observatory, P.O. Box 9, Observatory, 7935, South Africa
}

\date{\today}

% \abstract{}{}{}{}{} 
% 5 {} token are mandatory
 
\abstract
{Optimal frequency identification in astronomical datasets is crucial for variable star studies, exoplanet detection, and asteroseismology. Traditional period-finding methods often rely on specific parametric assumptions, employ binning procedures, or overlook the regression nature of the problem, limiting their applicability and precision.}
{We aim to introduce a universal, nonparametric kernel regression method for optimal frequency determination that is generalizable, efficient, and robust across various astronomical data types.}
{FINKER uses nonparametric kernel regression on folded datasets at different frequencies, selecting the optimal frequency by minimizing squared residuals. This technique inherently incorporates a weighting system that accounts for measurement uncertainties and facilitates multiband data analysis. We evaluate our method's performance across a range of frequencies pertinent to diverse data types and compare it with an established period-finding algorithm, conditional entropy.}
{The method demonstrates superior performance in accuracy and robustness compared to existing algorithms, requiring fewer observations to identify significant frequencies reliably. It exhibits resilience against noise and adapts well to datasets with varying complexity.}
{}


\maketitle

%-------------------------------------------------------------------

\section{Introduction}

In astronomy, periodic signals buried within time series of flux (photometric light curves) or radial velocity measurements serve as important carriers of scientific information. Their detection is essential for various scientific pursuits, from exoplanet characterization to the study of variable stars. A plethora of period-finding algorithms exist, from standard techniques such as the classical Lomb-Scargle periodogram \citep{Lomb1976,Scargle1982}, the generalized Lomb-Scargle \citep{Zechmeister2009,VanderPlas2018}, and the discrete Fourier transform \citep{Deeming1975} to popular nonparametric methods such as string length \citep{Dworetsky1983}, analysis of variance \citep{Schwarzenberg1989,Schwarzenberg1996}, and phase dispersion minimization \citep{Stellingwerf1978}, to more modern methods grounded in information theory such as conditional entropy \citep{Graham2013a}. In addition to standard periodogram algorithms, smoothing algorithms and statistical hypothesis testing based on regression models and confidence sets have also been used to detect periodicities in radial velocity measurements \citep{McDonald1986,Toulis2021}, highlighting the potential of regression techniques for periodicity detection in astronomical data. Furthermore, these methods can be used to search for periodicity either by grid search or direct optimisation \citep{Reimann1994}.

Recent comparative analyses have been instrumental in enhancing our understanding of various period-finding algorithms' capabilities and limitations \citep{Graham2013b}. Such studies emphasize the dependency of these methods on the quality of the light curve data. \citet{Graham2013b} advocates for a bimodal observation approach, where pairs of observations are taken rapidly each night to retain a high sampling frequency. They also emphasize that algorithms perform notably well for specific types of variables, such as pulsating and eclipsing classes, but found that ensemble methods \citep[e.g., ][]{Saha2017,Ranaivomanana2023}, which combine multiple algorithms, do not generally outperform single algorithms. Among the array of algorithms studied, conditional entropy stands out in terms of period recovery and computational time, with analysis of variance and phase dispersion minimization also showing promise.

Despite these advancements, the search for a universally applicable and assumption-free period-finding method continues. A noteworthy yet underdeveloped method presented by \citet{Hall2000} pioneered the exploration of nonparametric kernel regression techniques for period detection through grid search least-squares optimisation. Despite its potential, the method was not fully realized in astronomical applications due to the technological and computational limitations of the time. This work was later expanded in a series of papers to apply the framework of nonparametric regression estimation to periodograms using single and multiple sine/cosine components to circumvent the alias problem \citep{Hall2006,Genton2007}. This body of earlier work serves as a conceptual underpinning for our current research, while our approach substantially extends the methodology by developing a more advanced, nonparametric kernel regression algorithm tailored for astronomical applications. We further investigate the behaviour of the kernel bandwidth in frequency estimation.

Through this work, we aim to establish nonparametric kernel regression as a robust and versatile method for optimal frequency identification, setting the stage for more precise and efficient analyses in both astronomical and potentially interdisciplinary fields. The structure of this paper is as follows: Section \ref{Sec: Method} contains the mathematical foundations and computational aspects of our proposed method. Section \ref{Sec: Applications} describes the experimental setup and conducts a comparative analysis with existing methodologies. Section \ref{sec: application_real} presents the results of FINKER on a set of real examples, such as classical pulsators, short period transiting compact binaries, and radial velocity variations in binaries. Finally, Section \ref{Sec: Conclusion} offers conclusions and explores avenues for future developments.

To promote open and reproducible research, FINKER is publicly available on GitHub\footnote{\url{https://github.com/FiorenSt/FINKER}}.

\section{Method}
\label{Sec: Method}

This section details the methodology employed for nonparametric kernel regression based frequency optimization in astronomical data sets. We discuss the relevant data preprocessing, the technical implementation of the nonparametric regression model, and the determination of the optimal frequency of variability in the time series. Each subsection delves into the specifics of these steps, elucidating the underlying principles and computational strategies. In general, we consider an arbitrary, sparsely sampled astronomical time series, consisting of $N$ observations $y_i$ with corresponding uncertainties $\sigma_i$ taken at discrete times $t_i$ (with $i=1,...,N$). This pertains to both photometric flux and radial velocity time series.

\subsection{Phase folding}

\begin{figure}[h!]
    \centering
    \includegraphics[width=0.48\textwidth]{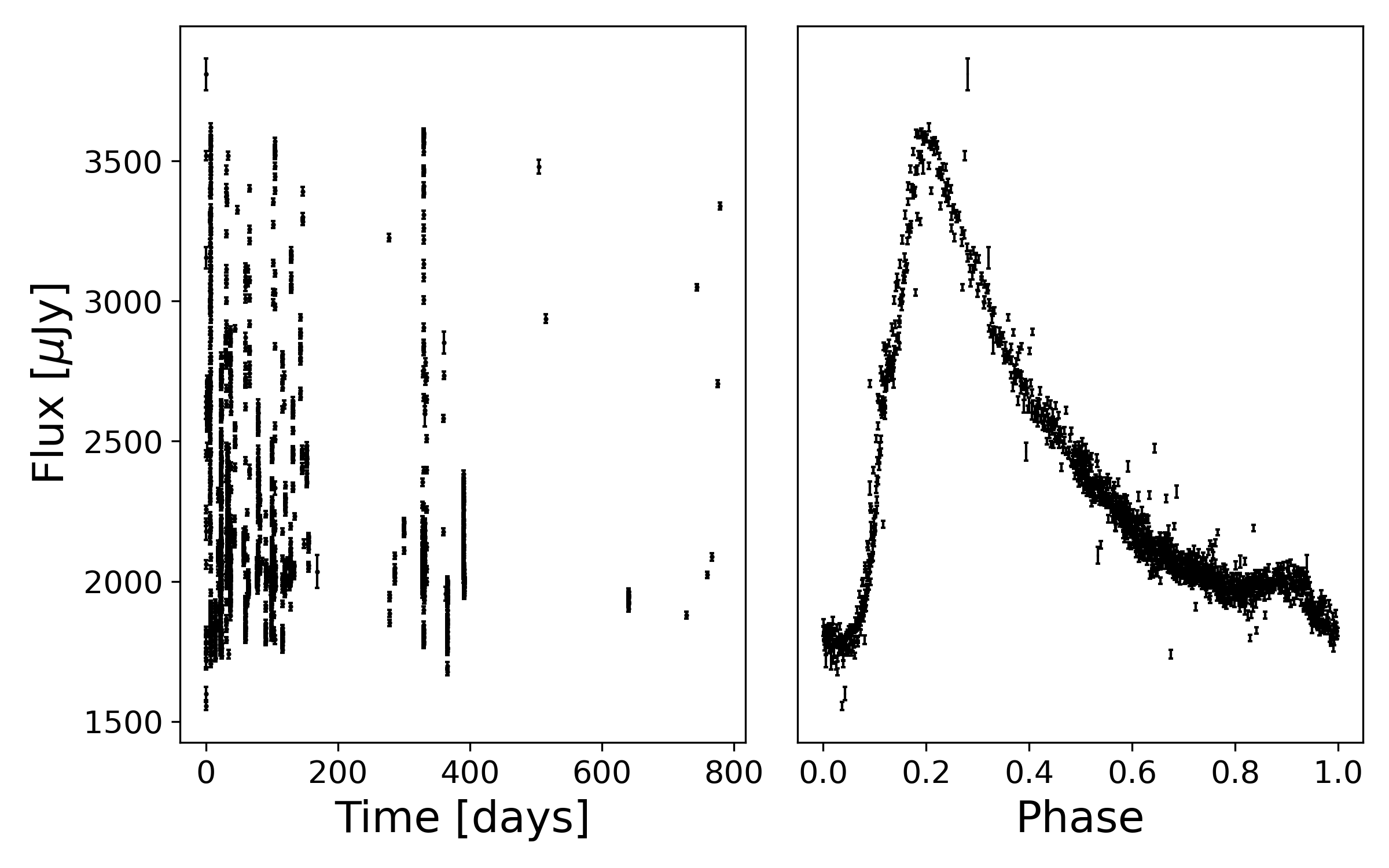}
    \caption{MeerLICHT \citep{Bloemen2016} telescope's light curve of CRTS J033427.7–271223 on the left and phase-folded version at the literature frequency \citep{Torrealba2015} on the right. Associate to each observation is an estimated measurement uncertainty.}
    \label{fig:folded_light_curve_source1}
\end{figure}

Phase folding is a common processing step in the analysis of periodic signals within astronomical data. It involves transforming the time-series observations into a phase diagram by folding the data over a chosen frequency $\nu_j$, such that the phase of a given observation $\phi_i$ is given by:

\begin{equation}
    \label{eq:phase}
    \phi_i = (t_i - t_0) \; \nu_j \mod 1, 
\end{equation}

\noindent
where $t_0$ is the chosen reference epoch\footnote{We note that the reference epoch refers to different moments for different types of data. For example, with eclipsing binaries and transiting systems $t_0$ is set to coincide with the time of superior conjunction, whereas in spectroscopic binaries $t_0$ refers to the time of periastron passage.}, and $\nu_j$ is the candidate frequency for folding. This technique aligns the repeating patterns in the data, which can be obscured in the time domain due to irregular sampling intervals or noise. The periodic signal becomes more discernible by folding the data at the correct or assumed frequency, facilitating the analysis of its structure and properties. The phase-folded curve is particularly useful for visualizing and analyzing periodic variations in the brightness of variable stars or exoplanets transiting their host stars and for radial velocity variations of multiple star systems or exoplanetary systems. To illustrate this, Figure \ref{fig:folded_light_curve_source1} showcases a real light curve alongside its phase-folded counterpart at the literature frequency \citep{Torrealba2015}. Upon examining this example, we notice that by folding the light curve on the underlying period of variability, we reduce the scatter between two adjacent points. This is the basis for various nonparametric frequency determination routines, including the Analysis of Variance method  \citep{Schwarzenberg1989}, the Lafler-Kinman statistic \citep{Clarke2002}, the conditional entropy periodogram \citep{Graham2013a}, and our proposed nonparametric kernel regression technique. 

\subsection{Nonparametric kernel kegression}

Nonparametric kernel regression \citep{Nadaraya1964,Watson1964,Hall2000} serves as the core of our approach. In the context of this example used to demonstrate the methodology, the regression is applied to the relationship between phase and flux for photometric time series. Unlike parametric methods that assume a functional form for the underlying data, nonparametric methods aim to construct an estimate that adapts to the local structure of the data points. Since two successive observations may be well separated in time and thus not hold meaningful information on the periodically correlated nature of the observations, we transform the observations to phase-space following Eq.~\ref{eq:phase} to perform kernel regression within the phase-flux relationship. Additionally, to address the common boundary effect problem that kernel regression usually faces, we use a common trick of extending the phase-folded light curve to the left and right. This extension effectively mitigates boundary issues, allowing for a more accurate estimation of the regression function near the edges.

Local constant regression is the simplest variant and most efficient version of kernel regression. The regression function $ \hat{f}_{\text{lc}}(\phi) $ at any given phase $ \phi $ calculated for a given proposed frequency $\nu_j$ is computed as:

\begin{equation}
\hat{f}_{\text{lc}}(\phi) = \frac{\sum_{i=1}^{n} K_h(\phi - \phi_i) \, y_i}{\sum_{i=1}^{n} K_h(\phi - \phi_i)}
\end{equation}

\noindent
Here, $ y_i $ represents the observed flux corresponding to the phase $ \phi_i $, and the estimated function value $ \hat{f}(\phi) $ is a weighted average of these fluxes. The weights are determined by the kernel function $ K_h $, which measures the closeness of each observed phase $ \phi_i $ to the phase $ \phi $ of interest. This approach is computationally straightforward and only requires one parameter, the bandwidth $h$ of the kernel $K_h$. Section \ref{sec: bandwidth} explores this parameter in detail.

A simple extension to local constant, local linear regression \citep{Fan1994}, enhances the local constant model by incorporating a linear trend within the kernel's scope at each phase $ \phi$. This method is particularly famous for its ability to better manage boundary problems, an advantage in typical scenarios. However, in our approach, this boundary issue is already effectively addressed by extending the phase-folded light curve to both sides. This variant, while offering a refined analysis by including linear trends, necessitates the calculation of two additional parameters at each phase: the intercept and slope of the local linear model. Given its considerable computational load and the fact that our extensive frequency folding requirements do not show marked improvements over local constant regression, we opt for the latter. The computational simplicity and efficiency of local constant regression make it the more suitable choice for our frequency optimization framework.

\subsection{Choice of kernel}

The array of kernel functions available to researchers is rich, with each kernel bringing its own unique advantages to different data challenges \citep{Epanechnikov1969, Gasser1985,Izenman1991}. Despite this variety, our investigation specifically leverages the Gaussian kernel. This decision is informed by the kernel's prevalent application across various fields and the observation that, for the problems at hand, it tends to provide results that are not significantly different from those obtained using other kernels. The Gaussian kernel is mathematically expressed as:

\begin{equation}
K(\phi, \phi', h) = \exp\left(-\frac{(\phi - \phi')^2}{2 \, h^2}\right)
\end{equation}

\noindent
In this equation, $ \phi $ and $ \phi' $ are points in the feature space, and $h$ is the bandwidth, or relevant length scale, and can be optimized for specific applications.

One of the primary advantages of the Gaussian kernel is its smoothness. Being infinitely differentiable, it ensures that the estimated function is smooth. This is particularly beneficial for capturing underlying trends in astronomical data, which are often smooth in nature. The Gaussian kernel also has the property of localized influence, assigning significant weight only to points that are close to the target point in the feature space. Overall, the Gaussian kernel is a versatile tool for a wide range of applications.

\subsection{Bandwidth selection}
\label{sec: bandwidth}

In kernel regression, the choice of bandwidth $h$ is critical, as it directly impacts the estimator's bias and variance. A well-chosen bandwidth balances the trade-off between the smoothness of the estimated function and the fidelity to the data points.

Scott's Rule \citep{Scott1979} and Silverman's Rule \citep{Silverman1987} are both prevalent methods for determining the bandwidth of a kernel density estimate. Scott's Rule, typically represented as 
\begin{equation}
h = 3.49 \, \sigma \, n^{-1/3},
\end{equation}
where $n$ is the sample size and $\sigma$ is the standard deviation of the observations. This bandwidth aims to minimize the mean integrated squared error (MISE) for data that approximates a normal distribution, yielding a smoother density estimate suitable for elucidating the data's overall structure.
Conversely, Silverman's Rule, which is often given by 
\begin{equation}
h = 1.06 \, \sigma \, n^{-1/5},
\end{equation}
employs a slightly different formula tailored for Gaussian-like data but allowing for a tighter bandwidth, enhancing the detection of finer structural details within the data distribution.

Both rules presuppose a normal distribution and might not be optimal for datasets that are multimodal or exhibit heavy-tailed distributions. 
Given the unique characteristics of our data, we adopt a custom bandwidth formula that better aligns with the non-Gaussian, periodic nature of such datasets

\begin{equation}
h = \alpha \; n^{-1 / 5}.
\end{equation}

\noindent
This formula is derived from the same principle of minimizing the MISE, where the $ n^{-1 / 5} $ term balances the estimator's bias and variance as the sample size $ n $ increases. The constant $\alpha$ replaces the standard deviation component from Silverman's and Scott's Rules, allowing for adjustment based on the specific characteristics of the periodic data under study. This approach does not assume a Gaussian distribution, making it more adaptable to the heteroskedasticity and periodicity inherent in astronomical time-series data. Section \ref{sec: optimal_bandwidth} explains in more depth how we found an empiric optimal $\alpha$ for folded lightcurves.

Additionally, we built and tested a more accurate yet computationally expensive adaptive bandwidth strategy that assigns a different bandwidth to each data point, similar to the methodology proposed in \citet{Terrell1992} and \citet{Orava2012}. The bandwidth is calculated as the average distance of each point to its $k$ nearest neighbours, allowing the model to adapt to different levels of sparseness in the data. 

\noindent
We have chosen the number of neighbours, $k$, as a function of the sample size, $n$, specifically $\ln(n)$. This choice of $k$ differs from the asymptotic value of $k=n^{4/5}$ suggested in \citet{Orava2012}; however, it allows us to adapt more effectively to varying data densities, especially for small datasets. In datasets with thousands of observations, this choice would lead to overfitting, however, at those sample sizes, the use of an adaptive bandwidth is anyway impeded by the computational cost.

\noindent
The algorithm used for effectively finding the nearest neighbours is the Ball Tree algorithm \citep{Omohundro2009}; however, having to repeat this operation for each folding makes the adaptive bandwidth method more computationally expensive than the custom (fixed) bandwidth.

%Additionally, we built and tested a more accurate yet computationally expensive adaptive bandwidth strategy that employs nearest-neighbour trees \citep{Friedman1976A} to adapt the bandwidth dynamically, similar to the methodology proposed in \citet{Terrell1992} and \citet{Orava2012}.
% The bandwidth for each data point is calculated using the average distance to its $ k $ nearest neighbours, allowing the model to adapt to different levels of sparseness in the data. The method can be represented as

% \begin{equation}
% h_i = \frac{1}{k} \sum_{j=1}^{k} \text{Distance to the $ j^{\text{th}} $ nearest neighbor of $ \phi_i $}
% \end{equation}

% \noindent
% Here, $ h_i $ is the bandwidth for the $ i^{\text{th}} $ data point $ \phi_i $, and $ k $ is chosen as $ \ln(n) $, where $ n $ is the sample size. This choice of $ k $ allows the model to adapt more effectively to varying data densities, especially for large datasets.

\subsection{Role of weighted residuals in frequency determination}

The sum of squared residuals ($ SSR $) is a commonly used metric for assessing the goodness of fit. In the specific context of this work, it serves as a specialized objective to compare how the fit of the kernel regression behaves for data folded at different periods. The central premise is that a lower $ SSR $ indicates a more compact, smoother folded light curve, suggesting a more accurate fit to the inherent periodicity in the data. The $SSR$ is defined as

\begin{equation}
SSR_j = \sum_{i=1}^{n} \left(y_i - \hat{f}(\phi_i; \nu_j)\right)^2,
\label{eq: residuals}
\end{equation}

\noindent
where, $ \hat{f}(\phi_i; \nu_j) $ represents the estimated value at phase $ \phi_i $ when the time-series data is folded at frequency $ \nu_j $, as obtained through a nonparametric regression model.

However, the standard $ SSR $ gives equal weight to all residuals, ignoring measurement uncertainties associated with the observations. To address this, we simply use a weighted sum of squared residuals, denoted $ SSR_w $, that incorporates these uncertainties.

\begin{equation}
SSR_w = \sum_{i=1}^{n} \frac{\left(y_i - \hat{f}(\phi_i; \nu_j)\right)^2}{\sigma_i^2}
\label{eq: weighted_residuals}
\end{equation}

\noindent
This weighted $ SSR_w $ accounts for the heterogeneity in observational data quality, making the frequency selection more robust.

\subsection{Uncertainty estimation in frequency determination via bootstrapping}

To enhance the reliability of FINKER, we employed a bootstrap methodology \citep{Efron1979, Efron1986} that provides an estimate of the uncertainty of the predicted frequency. This entails generating multiple datasets from the original by random sampling with replacement, after which FINKER is applied to a finely tuned grid centring on the initial frequency estimate. We use the standard deviation of the bootstrap results as an estimate of the prediction reliability.

\noindent
In Section \ref{sec: uncertainties_on_synthetic}, we validate this approach on a set of simulations and show its reliability, comparing it to the known true errors in frequency.

Although a valid and informative method, its results are only meaningful if a frequency value has already been found with a certain degree of accuracy. Suppose the frequency value has been wrongly estimated. In that case, the measure of its uncertainty will remain small due to the search in a localized space around the found minimum, but the resulting value will be meaningless. We do not yet have a solution for this eventuality but will be subject to more in dept statistical research in the future.

\subsection{FINKER's steps for optimal frequency identification}
\label{sec:algorithm}

The algorithm for optimization is outlined as follows:

\begin{enumerate}
    \item Generate a set of candidate frequencies $ \nu_1, \nu_2, \ldots, \nu_m $.
    \item For each candidate frequency $ \nu_j $:
    \begin{enumerate}
        \item Fold the time-series data on $ \nu_j $ according to Eq~\ref{eq:phase}.
        \item Apply nonparametric kernel regression on the folded data.
        \item Compute $ SSR_w $ using Eq. \ref{eq: weighted_residuals}.
    \end{enumerate}
    \item Identify the frequency $ \nu_{opt} $ that minimizes the $ SSR_w$
    \item Repeat B times:
    \begin{enumerate}
        \item Sample with replacement from the original dataset.
        \item Apply nonparametric kernel regression on a small range around $\nu_{opt}$.
        \item Identify the frequency $ \nu_{b} $ that minimizes the $ SSR_w$.
    \end{enumerate}
    \item Estimate the uncertainty of $\nu_{opt}$, $\sigma_{\nu}$, as the standard deviation of all the $\nu_b$s.

    % \[    f_{opt} = \arg\min_{\nu_j} SSR \]
\end{enumerate}

\noindent
By automating this sequence of operations, FINKER provides a reliable estimation of the intrinsic periodic nature of the data, focusing on the frequency that yields the minimum $SSR_w$.

\section{Application to synthetic data and benchmarking}
\label{Sec: Applications}

This section presents an empirical validation of FINKER and shows its performance on synthetic data mimicking various astronomical phenomena.

\subsection{Synthetic data}

The generation of synthetic light curves is a critical component of our simulation framework, allowing us to test the robustness and effectiveness of our frequency optimization algorithm under controlled conditions. These simulations are characterized by several adjustable parameters, each designed to replicate diverse observational scenarios and attributes of astronomical entities.

Firstly, we adjust the number of data points to emulate both sparsely and densely sampled observations, mirroring the variability encountered in astronomical data acquisition. The total duration of these observations is also configurable, setting the temporal extent for the generated light curves.

Central to our simulations is the emulation of the light curve's frequency components. We simulate the dominant periodic signal within the data through a primary frequency and its corresponding amplitude, which denotes the strength of this signal. Additionally, a secondary frequency and its amplitude are included to model objects exhibiting multiple periodic behaviours.
%Furthermore, our model incorporates parameters to simulate eclipse features, significant for studying eclipsing binary systems or transiting exoplanets. 

\noindent
%A baseline brightness level can be selected for the simulated astronomical object, representing its average brightness. Additionally, the flux of this object is correlated with the standard deviation of its observed brightness. This approach allows us to simulate the quality of observational data by incorporating a realistic level of photometric uncertainty, which is tailored to each baseline flux. This simulation aspect is informed by data obtained from the MeerLICHT telescope in Sutherland, South Africa \citep{Bloemen2016}, ensuring an accurate representation of observational conditions.

\begin{figure}[h]
    \centering
    \includegraphics[width=.48\textwidth]{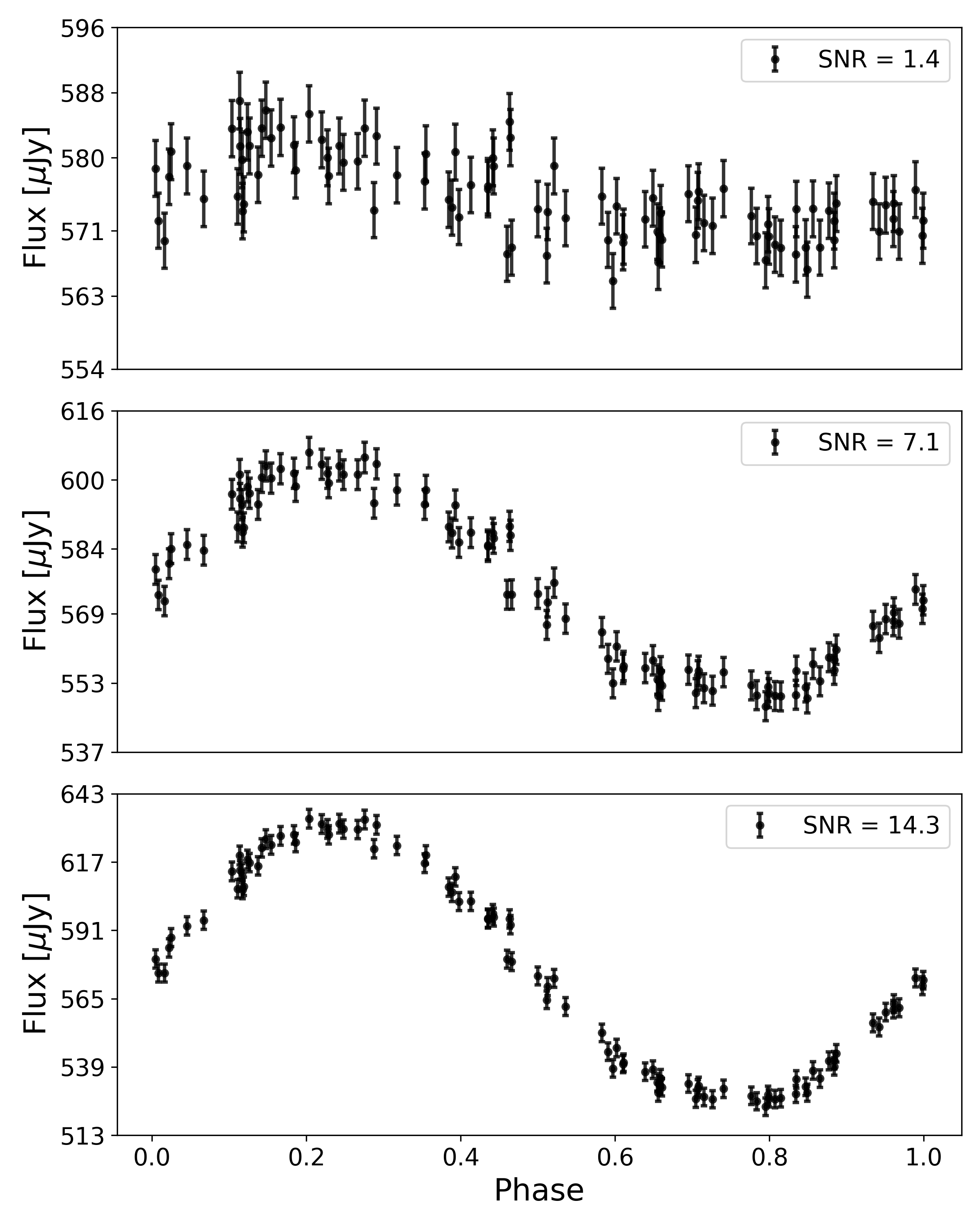}
    \caption{Simulated light curves at three SNR levels, based on the MeerLICHT telescope data-derived $\sigma$-magnitude function.}
    \label{fig:amplitudes}
\end{figure}

\noindent
A baseline brightness level can be selected for the simulated astronomical object, representing its average brightness. Additionally, we incorporate a realistic level of photometric uncertainty, informed by data obtained from the MeerLICHT telescope in Sutherland, South Africa \citep{Bloemen2016}. We built a function from real data that approximates the relation between the observations' uncertainty and their magnitude. So, for a fixed baseline magnitude, varying the amplitude is equivalent to having a varying amplitude-to-noise ratio. Figure \ref{fig:amplitudes} shows, for a baseline magnitude of 17, three levels of amplitudes and their significant variability in amplitude-to-noise. At m=17, we expect 0.007 mag, or $<1$\% scatter due to intrinsic noise. Therefore, the top, middle, and lower panels of Fig.~\ref{fig:amplitudes} correspond to the cases where the signal-to-noise ratio of the variation is 1.4, 7.1, and 14.3, respectively.

\subsubsection{Optimization of custom bandwidth parameter}
\label{sec: optimal_bandwidth}

As discussed in Section \ref{sec: bandwidth}, in our application, the only parameter influencing the performance of the kernel regression model is the kernel bandwidth $h$. 
As such, we analysed the effect of varying this parameter to find an optimal value that can be fixed for most types of light curves, and that would allow having zero parameters needed to be fine-tuned when using FINKER. 

Our proposed bandwidth, $h = \alpha \, n^{-1/5}$, is modulated by the multiplicative constant $\alpha$, which scales the base bandwidth $ n^{-1/5} $. 
The scaling factor based on the sample size $n$ allows to account for different light curve sizes, and, being the phase space always in the range [0,1], it also modulates the value of the bandwidth for different levels of $\delta_{\phi}$, the average distance between observations in phase. This, of course, assumes that in phase space, on average, observations are equidistant. In astronomical time series, this is not always the case, and for these scenarios, we propose a solution later on in this section.   

\begin{figure}[h]
    \centering
    \includegraphics[width=.48\textwidth]{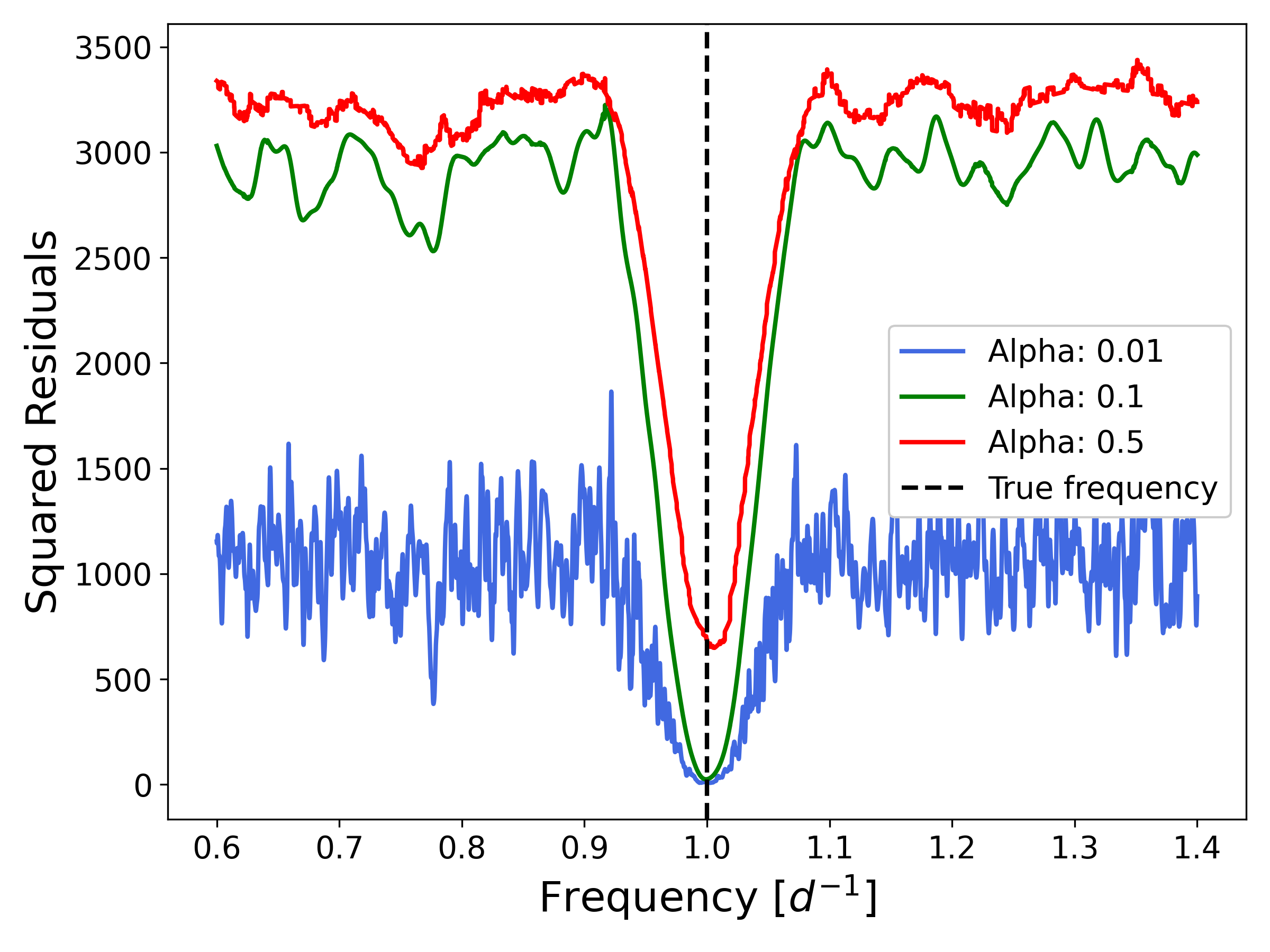}
    \caption{Folding frequencies and squared residuals for Kernel Regression with different $\alpha$ values. The vertical dashed line indicates the true frequency of 1 $d^{-1}$. All $\alpha$ values lead to frequency estimates that closely converge to the true value.}
    \label{fig:freq_reconstruction_alpha}
\end{figure}

\noindent
The other bandwidth component, $\alpha$, determines the smoothness of the regression estimate for a light curve folded at a specific frequency, thereby affecting the trade-off between overfitting and underfitting. 
Figure \ref{fig:freq_reconstruction_alpha} demonstrates the relationship between the folding frequencies and the residual sum of squares for three $\alpha$ values. An $\alpha$ value of $0.1$, around 10\% of the phase range ([0,1]), typically achieves a good balance. 

\noindent

Since the sample size is fixed in real scenarios and the phase space is always in the range $[0,1]$, we only need to find an $\alpha$ value that works in most situations. To do so, we built a set of simulations varying different signal-to-noise ratios (SNR) and sample sizes and searched for the alpha value that leads to the smallest estimated frequency error. We found that, on average, an $\alpha$ of $0.06$ will allow for an accurate frequency identification and consequentially remove the burden of hand-picking this parameter. This $\alpha$ value will be employed for the subsequent analyses in this study.

Our custom bandwidth approach is an adequate solution for a wide range of scenarios. However, in situations characterized by a limited number of observations, significant sparsity, or data concentrated in specific regions of the phase space, an adaptive bandwidth strategy emerges as a more effective alternative, enhancing accuracy in these particular conditions. This adaptive method, while significantly more computationally intensive, dynamically adjusts the bandwidth in response to local data density, thereby offering a more tailored fit to the underlying structure of the data. In Section \ref{sec: application_real}, we show how, for a select number of examples, an adaptive bandwidth allows to recover the optimal frequency with as small as ten observations.

\subsubsection{Bootstrap uncertainties reliability}
\label{sec: uncertainties_on_synthetic}

After having identified the best frequency, FINKER performs a bootstrap resampling of the original light curve and repeats the search process in a small grid around the best frequency. Repeating this process multiple times gives an estimate of the variability of the estimated frequency that can be used as an uncertainty estimate. 

In this section, we illustrate the application of FINKER's bootstrap-based uncertainty estimation method on a synthetic light curve. We demonstrate the method's potential in providing an uncertainty estimate around the determined frequency and visually assess the nature of these uncertainties in relation to various sample sizes and SNR.

\begin{figure}[h]
\centering
\includegraphics[width=0.49\textwidth]{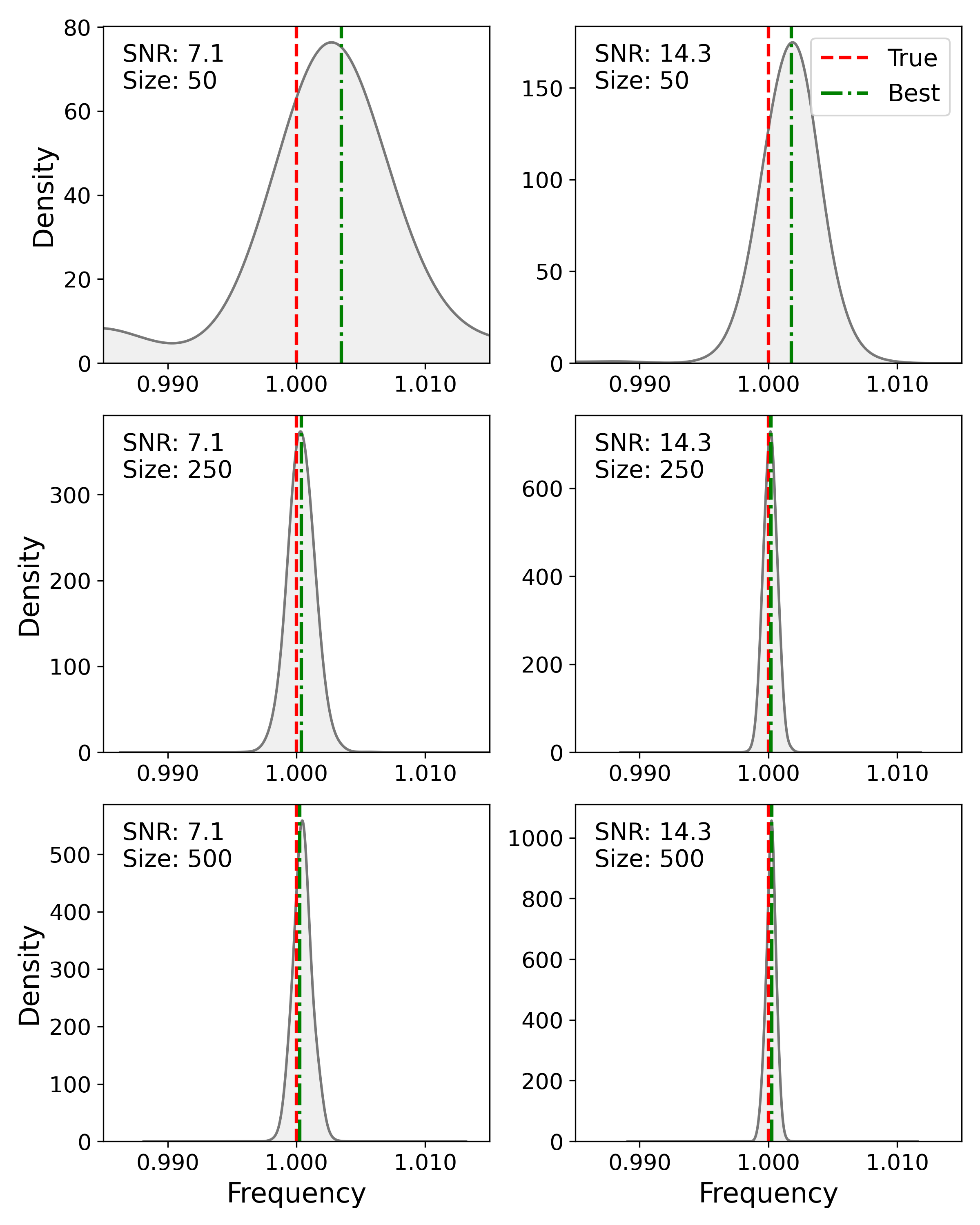}
\caption{Kernel density estimations (KDE) of bootstrap-estimated frequencies for a single synthetic light curve across varying sample sizes and SNR levels. Each subplot represents the bootstrap frequency distribution of the estimated best frequency juxtaposed with the true frequency of the synthetic light curve. As expected, at higher sample sizes and SNR, FINKER is more accurate, and its bootstrap density better resembles a Gaussian. Consequentially, the estimated frequency uncertainty is more reliable.}
\label{fig:errors_distribution}
\end{figure}

Figure \ref{fig:errors_distribution} presents a series of kernel density estimations (KDE) of the frequencies obtained through the bootstrap method for different combinations of sample size and SNR for a synthetic light curve. The alignment of the best frequency with the bootstrap distribution's mean suggests an unbiased nature of the estimation process. Notably, the bootstrap distribution deviates from the expected Gaussian shape at an SNR=7.1 with a sample size of 50, indicating that low-SNR signals or a small sample size may pose challenges to the reliability of the bootstrap method.

The standard deviation of the bootstrap frequencies, serving as the uncertainty measure, appears to offer a conservative estimate. This conservatism ensures that the estimated uncertainty is not understated, yet it also hints at the opportunity to refine the method to more accurately capture the true variation. Future iterations of FINKER will focus on improving the uncertainty estimation to ensure a more precise reflection of the underlying variability, particularly in challenging observational scenarios characterized by low SNR and limited sample sizes.

\subsubsection{Optimal frequency identification}

\begin{figure*}[]
\centering
\includegraphics[width=.95\textwidth]{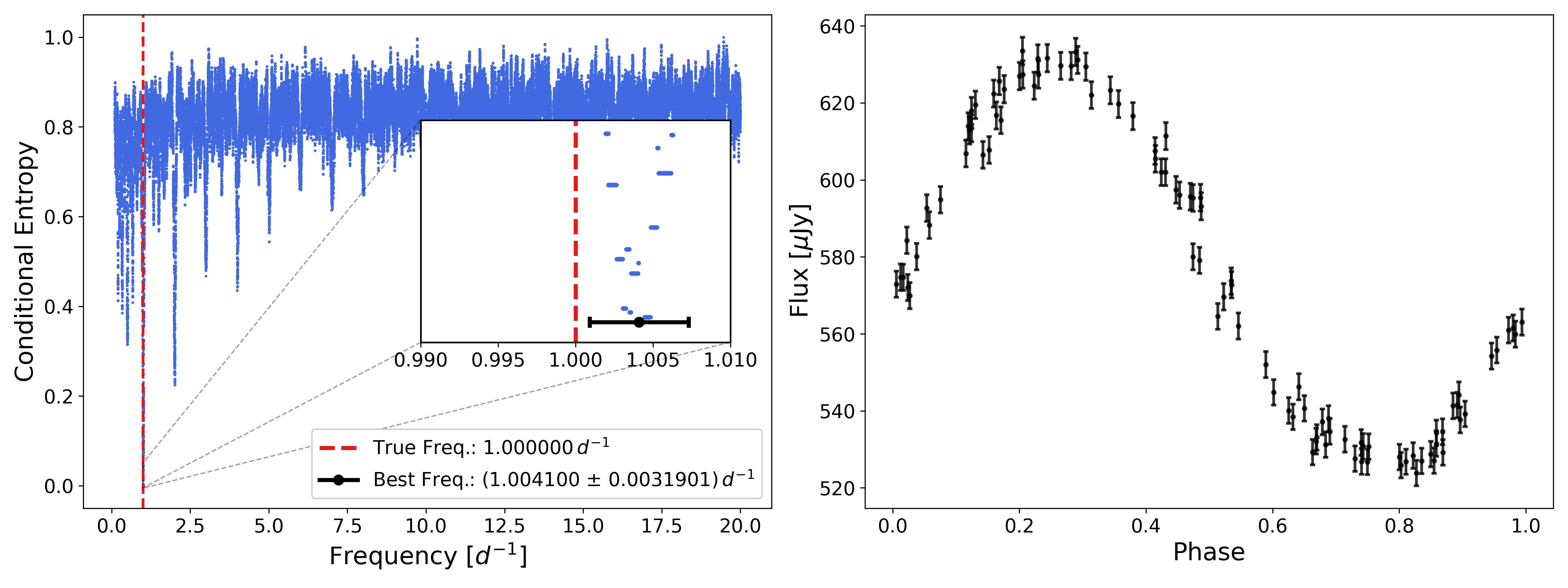}
\caption{Conditional entropy's results across a range of frequencies for a sinusoidal lightcurve with 100 observations, a baseline magnitude of 17, and an amplitude of 0.1 (SNR=14.3).  The plot on the left shows the frequency range searched, and the inset at the minimum entropy shows the noisy behaviour of the estimator. The estimated best frequency and its uncertainty are shown with a black error bar. The plot on the right shows the light curve folded at the found frequency.}
\label{fig: freq_reconstruction_entropy}
\end{figure*}

\begin{figure*}[]
\centering
\includegraphics[width=.95\textwidth]{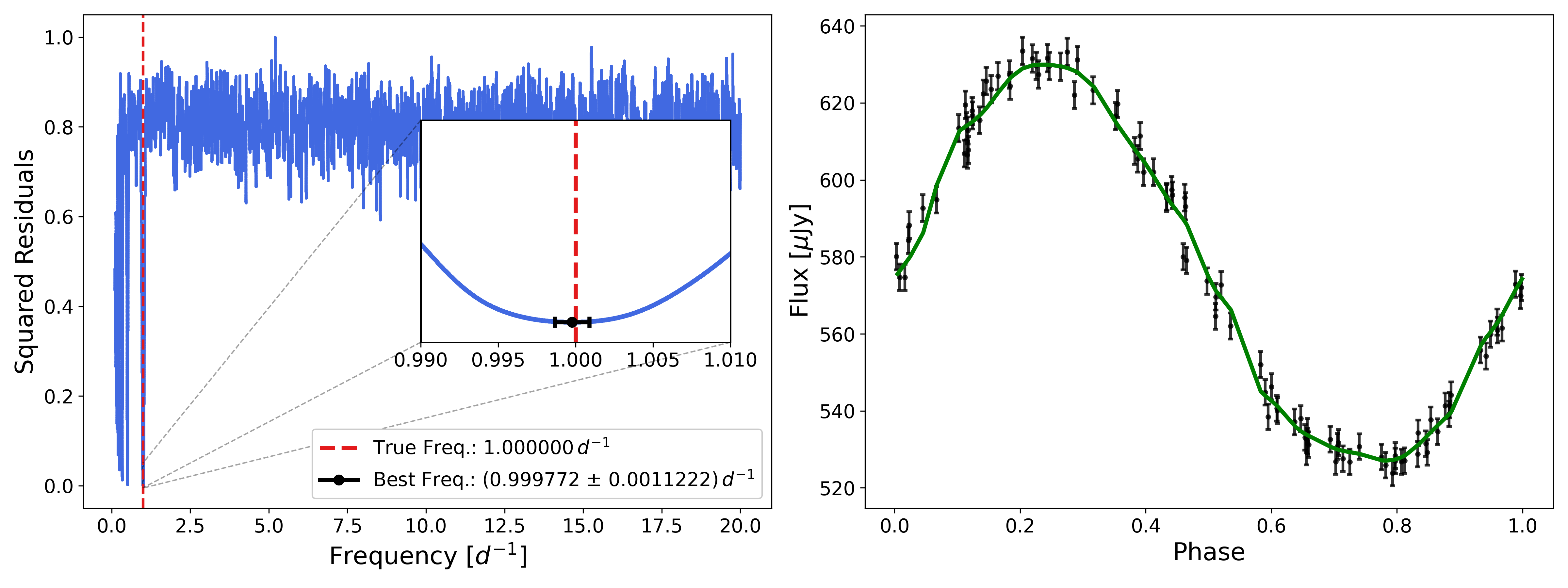}
\caption{FINKER's results across a range of frequencies for a sinusoidal lightcurve with 100 observations, a baseline magnitude of 17, and an amplitude of 0.1 (SNR=14.3). The plot on the left shows the frequency range searched, and the inset illustrates the smooth behaviour of the residuals around the minimum. The estimated best frequency and its uncertainty are shown with a black error bar. The plot on the right shows the light curve folded at the found frequency with the kernel regression fit (green line) overlaid, illustrating the algorithm's capability to model the periodic signal accurately.}
\label{fig: freq_reconstruction_kernel}
\end{figure*}

To validate the accuracy of FINKER, we conduct experiments using synthetic light curves with known true frequencies. Our kernel regression method is benchmarked against the commonly used conditional entropy \citep{Graham2013a}, which is employed for comparative analysis.

In short, conditional entropy gauges the variability of one variable, say the light curve's flux, given that we have knowledge of another, such as the phase. For a specific folding frequency, a low conditional entropy implies a reduced level of sparseness in the flux for a given phase, consequentially hinting that the frequency value chosen is correct. 

\noindent
In the computation of conditional entropy, a binning process via partition schemes is employed. This involves organizing the data into distinct bins or partitions, which aids in the accurate estimation of the probability distributions involved in the formula. Particularly, a simple rectangular partitioning scheme, with $i=1,...,N$ bins in flux space and $j=1,...,M$ bins in phase space is commonly adopted for computational efficiency during the analysis. The conditional entropy $ H(m | \, \phi) $ is defined as

\begin{equation}
H(m | \, \phi) = \sum_{i, \, j} p(y_i, \, \phi_j) \ln \frac{p(\phi_j)}{p(y_i , \, \phi_j)}.
\end{equation}

\noindent
Here, $ p(y_i, \phi_j) $ denotes the estimated probability of a data point occupying the $i$th flux bin and the $j$th phase bin simultaneously, while $p(\phi_j)$ signifies the estimated probability of a data point falling within the $ j $th phase bin, irrespective of its flux.
In the context of analysing light curves, conditional entropy is used to infer the frequency value that minimizes this entropy. Algorithmically, frequency identification using conditional entropy follows a similar approach to what is laid out in Section~\ref{sec:algorithm}, but with a different objective function. We fixed the binning size for all subsequent analyses to $N=10$ and $M=10$.

Figures \ref{fig: freq_reconstruction_entropy} and \ref{fig: freq_reconstruction_kernel} illustrate the results of the conditional entropy and FINKER, respectively, for the search of the optimal frequency on a sinusoidal light curve. The left figure shows the range of frequencies tested and the objective function values for each method, the SSR for ours and conditional entropy for the benchmark method. The two objective values are not directly comparable, but for visual clarity, we rescaled both metrics in a $[0,1]$ range for all subsequent plots. The right figure shows the light curve folded at the estimated best frequency and, for FINKER, the associated kernel regression fit.

As shown in Fig. \ref{fig: freq_reconstruction_kernel}, FINKER yields an accurate frequency prediction and has a smooth behaviour in its estimator. The noisy results of the conditional entropy are instead likely attributable to the binning process needed for the calculation of its metric. 

\begin{figure}[h!]
    \centering
    \includegraphics[width=.48\textwidth]{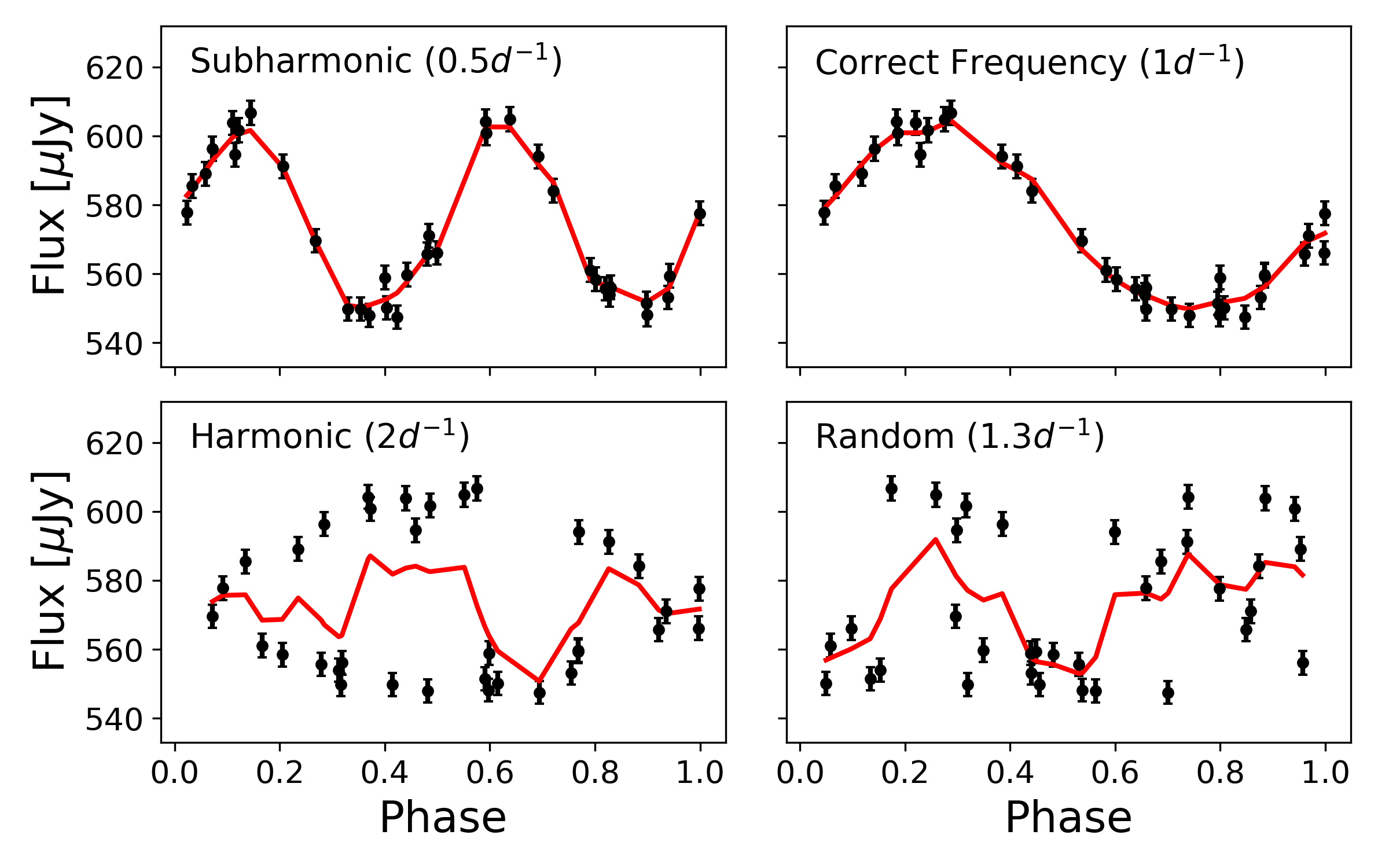}
    \caption{Kernel regression's fits (red lines) for harmonics and subharmonics of the true frequency. The light curve is folded at four different frequencies: half the true frequency (top left), the true frequency (top right), twice the true frequency (bottom left), and a random frequency (bottom right). Conveniently, for multiples of the true frequency, kernel regression will badly fit the observations resulting in high residuals.}
    \label{fig: harmonics}
\end{figure}

\noindent
Furthermore, our methodology inherently mitigates the common pitfall of mistaking harmonic frequencies for the fundamental frequency—a frequent issue with both other nonparametric approaches and standard parametric approaches. Specifically, when a frequency is an integer multiple of the true frequency, the light curve maintains a semblance of order and, consequentially, a relatively low entropy. Kernel regression, however, evaluates the goodness-of-fit for a frequency by analyzing the residuals post-smoothing. Harmonic frequencies, which introduce regular oscillations, are neutralized by the smoothing process, leading to high residuals similar to those from an incorrect frequency. This unique attribute of kernel regression allows it to identify the true frequency more accurately. This behaviour is demonstrated in Fig. \ref{fig: harmonics}.

\noindent
Despite its robustness against harmonic frequencies, kernel regression seems more sensitive to subharmonic frequencies with respect to the conditional entropy method. However, these are usually not misidentified since additional periodicity in the folded light curve will have an overly smoothed estimator and, consequentially, a higher residual. Furthermore, to counter this behaviour, our implementation includes a verification step to examine whether twice the identified optimal frequency yields a significant result. If the doubled frequency also presents a low residual sum of squares, it suggests that the true fundamental frequency may indeed be twice the initially identified one. This additional check enhances the method's accuracy and is a standard option in our code.

Overall, this additional property of kernel regression not only enhances its reliability in frequency estimation but also reduces the likelihood of the need for manual inspection or secondary validation methods to confirm the fundamental frequency. This robustness is particularly valuable in automated analysis pipelines where large volumes of data preclude detailed individual review.

\subsubsection{Sensitivity analysis at low sample size}

To rigorously assess the robustness and performance of our frequency optimization algorithm, a sensitivity analysis is conducted. This analysis focuses on two key parameters: the number of data points ($ n_{\text{points}} $) and the amplitude of the primary frequency of variability. These parameters are systematically varied over a predefined range to simulate different observational conditions. For each combination of $ n_{\text{points}} $ and amplitude, the algorithm is executed using two different methods: kernel regression with custom bandwidth and conditional entropy. All simulations are calculated using the same baseline magnitude (17), such that changing the amplitude of the variability will result in different signal-to-noise levels.

\begin{figure}[h]
    \centering
    \includegraphics[width=0.48\textwidth]{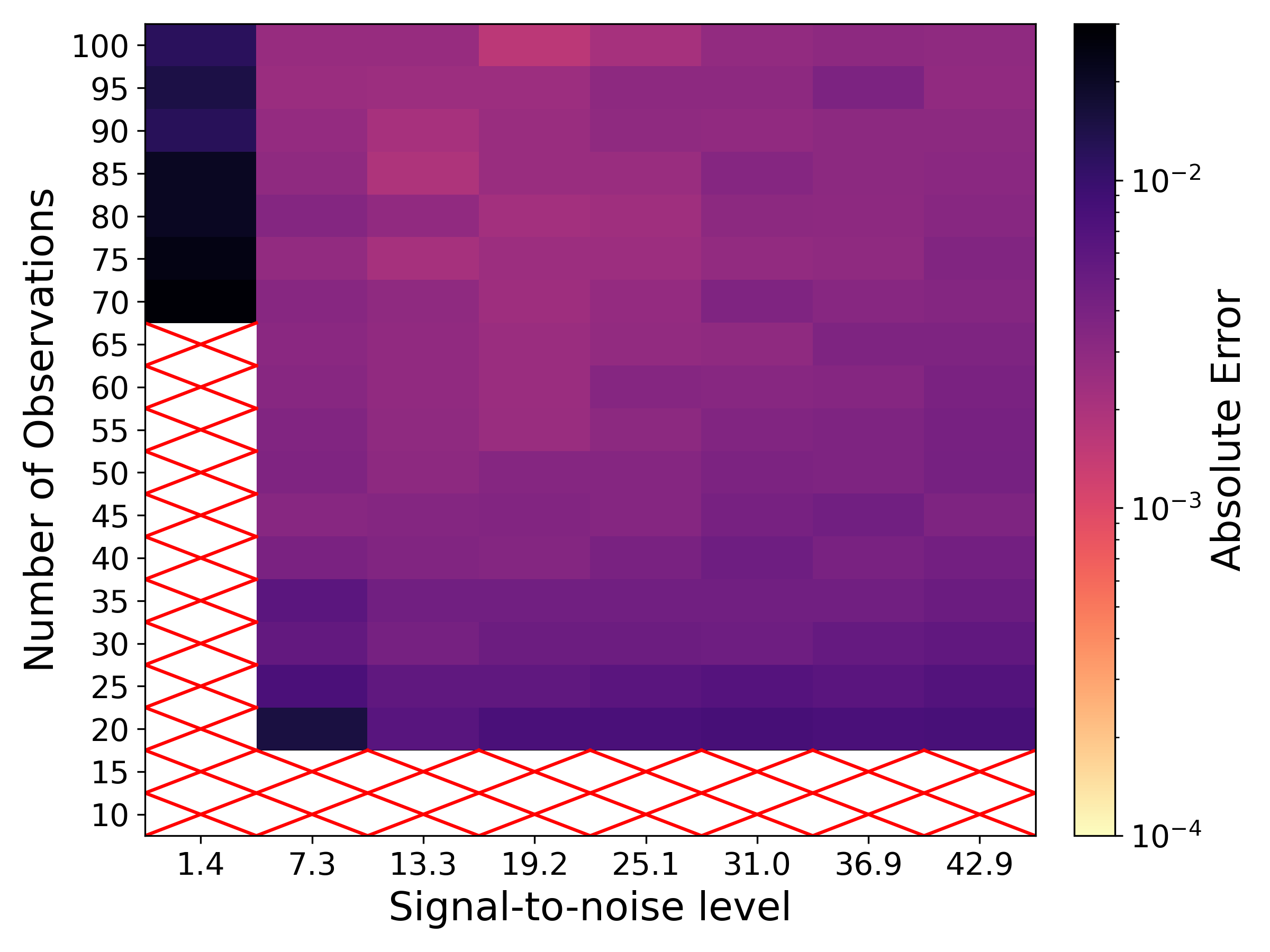}
    \caption{Heatmap of the Conditional Entropy sensitivity analysis, illustrating the absolute error in frequency estimation relative to sample size (y-axis) and amplitude (x-axis). Lighter shades correspond to lower absolute errors, with cells marked by a red cross representing an average absolute error above 0.03, deemed as unreliable estimates. This analysis indicates that at least 20 observations are necessary for reliable frequency estimation at most amplitude levels, excluding the lowest one.}
    \label{fig:heatmap_entropy}
\end{figure}

\begin{figure}[h]
    \centering
    \includegraphics[width=0.48\textwidth]{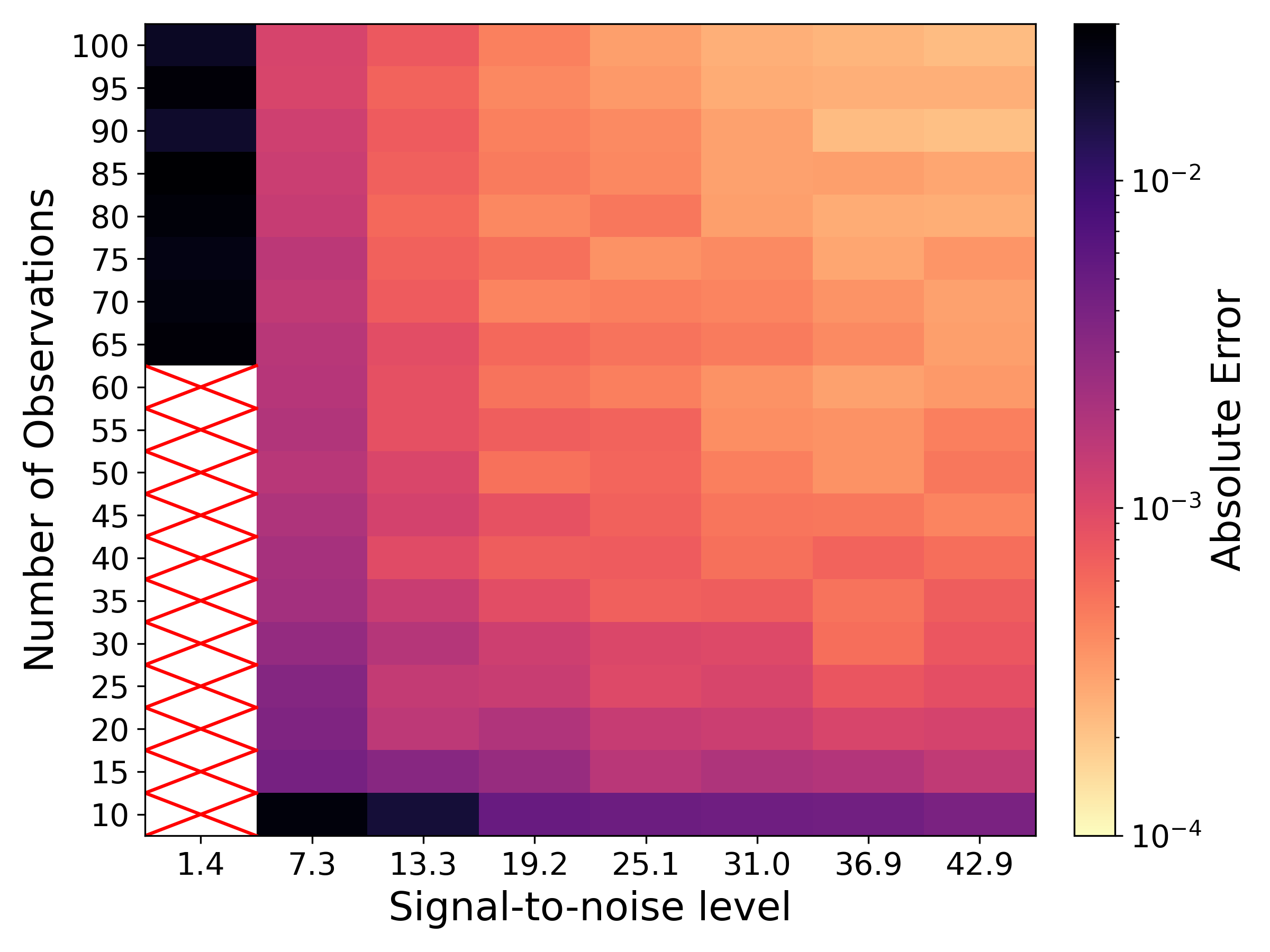}
    \caption{Heatmap of the Kernel Regression sensitivity analysis, depicting the absolute error in frequency estimation as a function of sample size (y-axis) and amplitude (x-axis). Lighter shades correspond to lower absolute errors, with cells marked by a red cross representing an average absolute error above 0.03, deemed as unreliable estimates. The heatmap suggests that approximately 10 observations are sufficient for a reliable frequency estimate across various amplitude levels, with the only exception being the lowest amplitude level.}
    \label{fig:heatmap_kernel}
\end{figure}

The results are visualized for our kernel regression periodogram and the conditional entropy periodogram in Fig. \ref{fig:heatmap_entropy} and \ref{fig:heatmap_kernel}, respectively. For both figures, the x-axis represents the amplitude levels, and the y-axis represents the number of observations simulated. The colour intensity on each heatmap indicates the absolute error between the estimated and true frequency (of 1 $d^{-1}$), $|f_{true}-f_{pred}|$.

\noindent
The heatmaps reveal a distinct contrast in performance between the methods. For a sinusoidal light curve, our kernel regression method demonstrates remarkable accuracy, with a clear trend of improvement as sample size and amplitude increase. In contrast, the conditional entropy method exhibits a dependency on larger data sets to attain comparable levels of accuracy, requiring at least twofold the number of observations to match the accuracy of our kernel regression technique. This robustness is critical for practical applications with scarce and sparse observational data.

\section{Application to real data}
\label{sec: application_real}

The efficacy of our frequency optimization algorithm is further substantiated through its application to a diverse array of real astronomical datasets. These datasets encompass light curves from variable stars, radial velocity measurements, and photometric data of transiting exoplanets. The sources of these datasets are diverse, originating from various telescopes and surveys, including the MeerLICHT \citep{Bloemen2016} and Zwicky Transient Factory (ZTF, \citealp{Bellm2019}) photometric missions and the HERMES spectrograph \citep{Raskin2011}. 

\subsection{Photometric time series}

MeerLICHT is a fully robotic telescope located in Sutherland Observatory in South Africa and is equipped with five Sloan $ugriz$ photometric filters with an additional custom $q$ filter that is a combination of the $g$ and $r$ filters \citep{Bloemen2016}. MeerLICHT nominally integrates for 60 seconds and has several observing strategies aimed at identifying and characterising transients in multiple wavelengths. The combination of these observing strategies has resulted in a large database of heterogeneously sampled, (sometimes) contemporaneous multi-colour photometric time series of millions of objects compiled over the mission five year runtime. As the MeerLICHT light curves are highly irregularly sampled, they are ideal candidates for testing the performance of our algorithm.

Prior to analysis, the MeerLICHT data is subjected to a standard preprocessing pipeline. This includes the removal of outliers, correction for atmospheric extinction, and normalization to account for instrumental variations (we refer the readers to \citet{deWet2021} and \citet{Ranaivomanana2023} for a complete description of the data reduction process). 

In addition to MeerLICHT data, we also make use of ZTF data in the case of compact transiting binaries. ZTF is a robotic telescope equipped with three photometric filters that surveys the night sky for transients and periodic variable stars with 30 second exposures in each filter \citep{Bellm2019}. While the majority of the data is obtained in the $g$ and $r$ filters, we only consider data in the $g$ filter for this work. Due to its observing strategy, these data are irregularly sampled, but are less sparsely sampled than the MeerLICHT data. Additionally, ZTF has occasional deep drilling campaigns which cover select fields with a high cadence over a short period of time. Thus, the ZTF data has highly irregular sampling spanning several years.

Table \ref{tab:table_comparison} lists a set of known light curves and their literature frequencies. We also show frequencies and uncertainties determined by the conditional entropy and nonparametric kernel regression periodograms. Below is some more information on the type of light curves shown.

\begin{table*}[h]
    \centering
    \begin{tabular}{l|c|c|c|c|c|c}
         Target & Type & Conditional & Kernel     & Literature & Sample      & Mag\\
                &      & entropy     & regression & frequency &  size        & \\
                &      & [d$^{-1}$]    & [d$^{-1}$]   & [d$^{-1}$]  &              &  [mag]   \\
        \hline
         CRTS J033427.7--271223 & Clas. Pulsator & $1.717876$ $\pm$ $4.4 \, 10^{-5}$& $1.717831$ $\pm$ $6.0 \, 10^{-6}$& $1.717829^a$ & 1414 &  V=15.5 \\
         CRTS J033500.6--272854 & Clas. Pulsator & $1.433518$ $\pm$ $1.0 \, 10^{-5}$ & $1.433525$ $\pm$ $3.4 \, 10^{-6}$ & $1.433527^a$ & 1534 & V=14.7 \\
         MLT J033147.60--281307.9 & Eclips. Binary & $5.277603$ $\pm$ $1.1 \, 10^{-5}$ & $2.638751$ $\pm$ $4.9 \, 10^{-6}$ & $2.63875^*$ & 1350 & V=14.0 \\
         MLT J162036.64--614110.5 & Eclips. Binary & $3.754604$ $\pm$ $1.5 \; 10^{-5}$ & $3.754548$ $\pm$ $8.6 \; 10^{-6}$ & $1.87726^*$ & 808 & g=15.0 \\
         ZTF J041016.82--083419.5 & Comp. Binary & $12.329054$ $\pm$  $3.2 \, 10^{-5}$ & $12.329054$ $\pm$ $1.7 \, 10^{-5}$   & $12.329042^b$ & 352 & G=17.4 \\
         ZTF J053708.26--245014.6 & Comp. Binary & $10.985491$ $\pm$ $3.4 \, 10^{-5}$ & $3.050674^{\dagger}$ $\pm$ $1.0 \, 10^{-5}$ & $3.050700^b$  & 138 & G=16.7 \\
         ZTF J063808.71+091027.4 & Comp. Binary & $1.520599$ $\pm$ $8.1 \, 10^{-6}$ & $1.520579$ $\pm$ $5.1 \, 10^{-6}$ & $1.520576^b$  & 735 & G=19.0 \\
         ZTF J140702.57+211559.7 & Comp. Binary & $6.979319$ $\pm$ $1.3 \, 10^{-5}$ & $6.979318$ $\pm$  $5.9 \, 10^{-6}$ & $6.979331^b$  & 366 & G=18.1 \\
         HD 165246 & RV Binary & $0.217809$ $\pm$ $8.6 \, 10^{-4}$ & $0.217754$ $\pm$ $8.7 \, 10^{-5}$& $0.217737^c$& 95 & V=7.6 \\
         V772 Cas & RV Binary & $0.144847$ $\pm$ $3.7 \, 10^{-4 \; \ddagger}$ & $0.188593^{\dagger}$ $\pm$ $1.5 \, 10^{-2 \; \ddagger}$ & $0.199453^d$ & 10 & V=6.7 \\
         HD 114520 & RV Binary & $0.002284$ $\pm$ $7.4 \, 10^{-6}$ & $0.002284$ $\pm$ $5.5 \, 10^{-6}$ & $0.002283^e$ & 71 & V=6.8 \\
         % KIC 5217733 & RV Binary & 0.015169 $\pm$ & 0.011465$^{\dagger}$ $\pm$ & $0.006201^f$ & 81 & V=7.3 \\

    \end{tabular}
    \caption{Table comparing the optimal frequencies determined for each target using both the conditional entropy and kernel regression periodograms. (a) \citet{Torrealba2015}, (b) \citet{Brown2023} , (c) \citet{Johnston2021}, (d) \citet{Kochukhov2021}, (e) \citet{Escorza2019}, (*) frequency verified visually, ($\dagger$) obtained with FINKER's adaptive bandwidth, ($\ddagger$) unreliable uncertainty estimate due to low sample size.}
    \label{tab:table_comparison}
\end{table*}

\paragraph{\bf Classical pulsators: }

\begin{figure*}[h!]

        \centering
        \includegraphics[width=.95\textwidth]{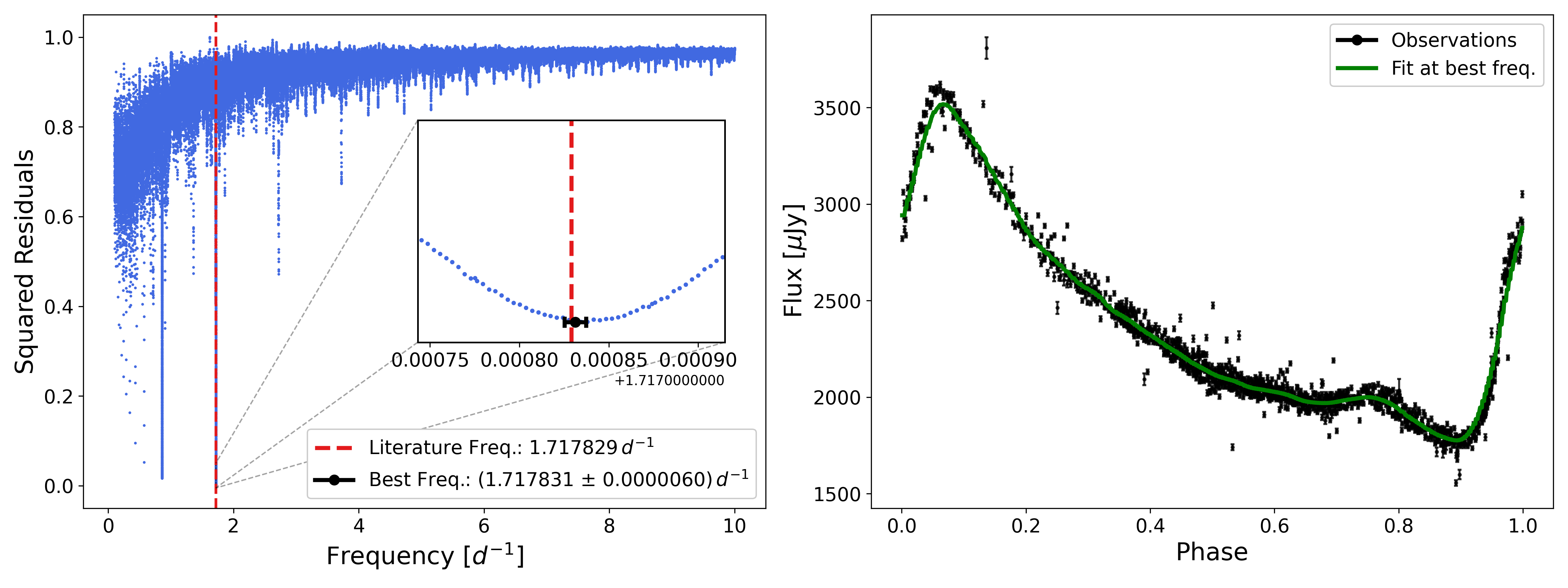}
        \caption{FINKER's frequency search for CRTS J033427.7--271223. The figure on the left shows the frequency range searched, and the inset illustrates the smooth behaviour of the residuals around the minimum. The estimated best frequency and its uncertainty are shown with a black error bar. The figure on the right shows the light curve folded at the found frequency with the kernel regression fit (green line) overlaid, illustrating the algorithm's capability to model the periodic signal accurately.}
        \label{fig:kernel_regression_rrlyra}
\end{figure*}

Classical pulsators are large amplitude, radially pulsating stars that exist in the classical instability strip, including $\delta$ Scuti variables, RR~Lyrae variables, and Cepheid variables \citep{Aerts2010,Kurtz2022}. We tested our algorithm on two classical pulsators from the MeerLICHT data and benchmarked it against the conditional entropy periodogram. In both cases, FINKER finds a similar optimal frequency as the conditional entropy periodogram, with a smaller uncertainty and seemingly closer to the literature value. We plot the kernel regression periodogram for one classical pulsator target (CRTS J033427.7--271223) in Figure \ref{fig:kernel_regression_rrlyra}.

\paragraph{\bf Eclipsing binaries: }

\begin{figure*}[h]
    \centering
    \includegraphics[width=.95\textwidth]{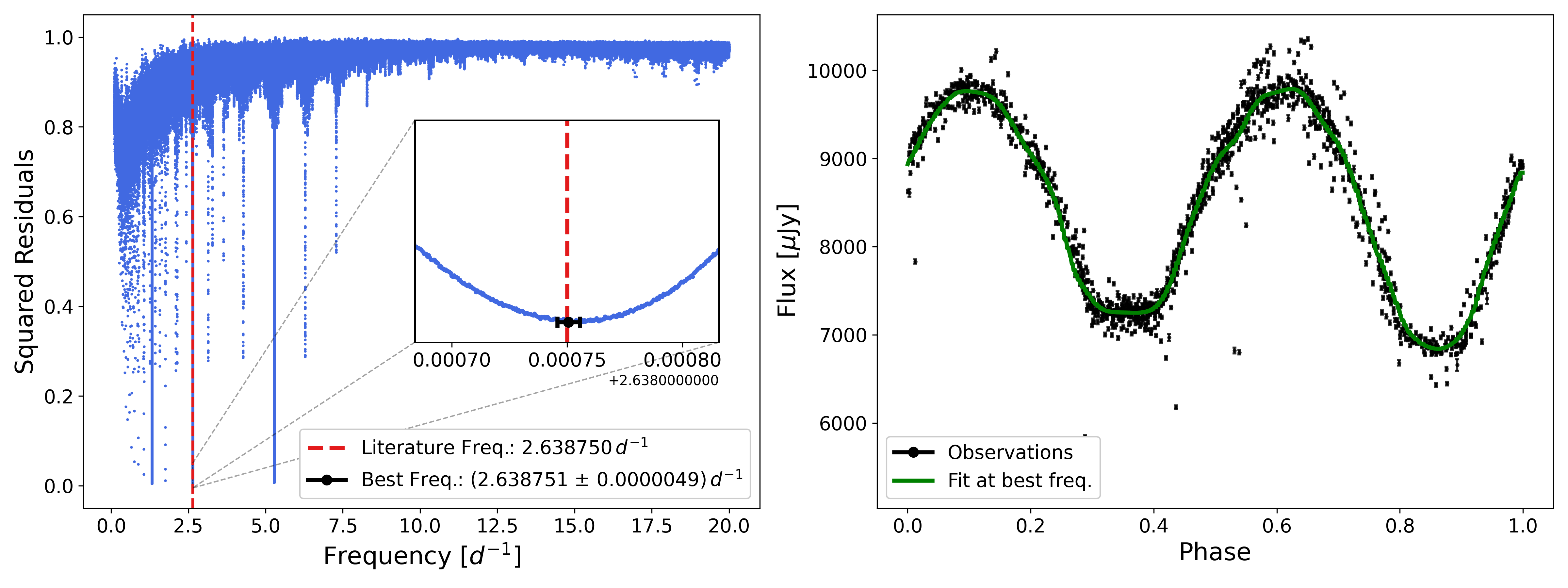}
    \caption{FINKER's frequency search for MLT J033147.60--281307.9. The figure on the left shows the frequency range searched, and the inset illustrates the smooth behaviour of the residuals around the minimum. The estimated best frequency and its uncertainty are shown with a black error bar. FINKER's framework allowed for a search of both the harmonic and subharmonic of the best frequency and correctly identified that there is a difference in the eclipses. The figure on the right shows the light curve folded at the found frequency with the kernel regression fit (green line) overlaid, illustrating the algorithm's capability to model the eclipsing signal accurately.}
    \label{fig:kernel_regression_EB}
\end{figure*}

The majority of stars exist in binaries or higher order multiples \citep{Moe2017, Offner2023}. If the orbital plane is inclined favourably with respect to us, we can see the stars eclipse one another as they move through their orbit. This produces periodic decreases in light that can range from less than 1\% to blocking nearly all of the light from one star. As most stars are in binaries, eclipsing binaries are commonly found in photometric time series. Here, we look at two eclipsing binaries to demonstrate our method's ability to reliably identify binaries with more complicated phase behaviour than classical pulsators and sinusoidal variables. Figure \ref{fig:kernel_regression_EB} illustrates the kernel regression periodograms for the eclipsing binary MLT J033147.60--281307.9.

\paragraph{\bf Short period transiting compact binaries: }

\begin{figure*}[h]
    \centering
    \includegraphics[width=.95\textwidth]{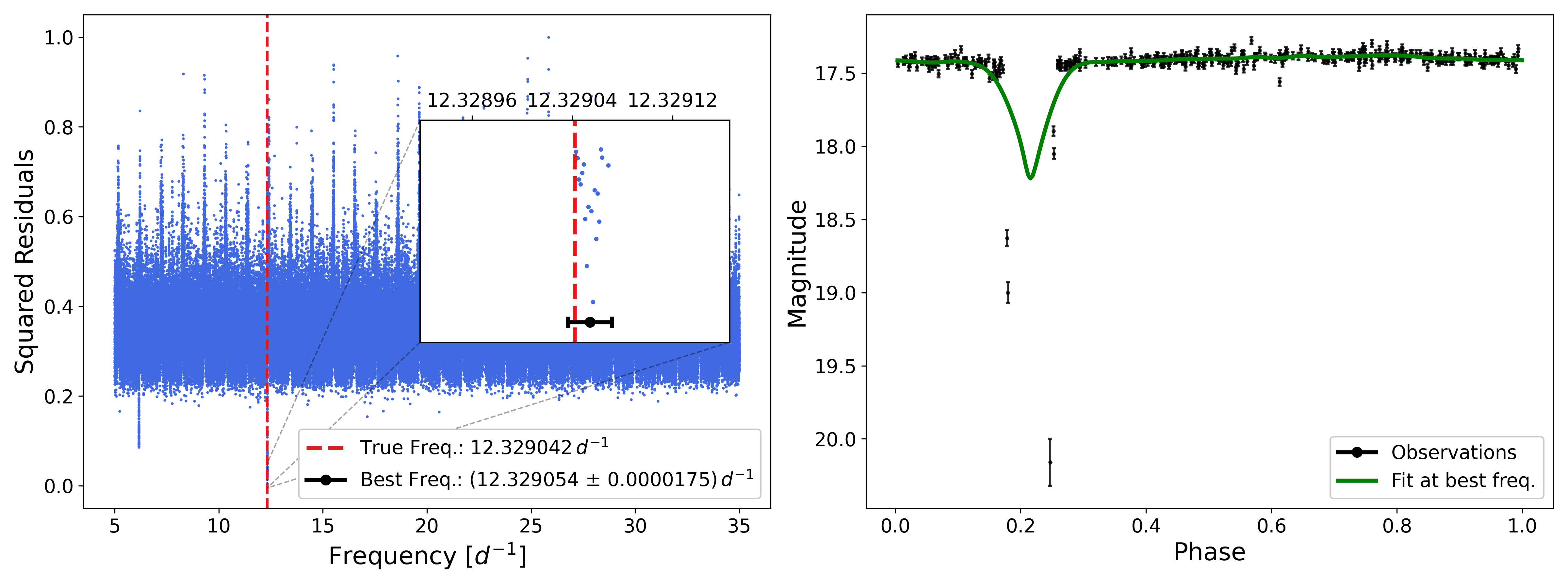}
    \caption{FINKER's frequency search for ZTF J041016.82--083419.5. The figure on the left shows the frequency range searched, and the inset illustrates the behaviour of the residuals around the minimum. Transit data shows a less smooth behaviour of the residuals around the minimum, but the frequency and its uncertainty, shown as a black error bar, are consistent with the literature value. The figure on the right shows the light curve folded at the found frequency with the kernel regression fit (green line) overlaid.}
    \label{fig:kernen_regression_AMCV2}
\end{figure*}

Ultra short period binaries that contain at least one compact component (e.g. a white dwarf) are important contributors to gravitational wave events and exotic transient phenomena. In particular, we consider binaries where a white dwarf and M-dwarf star orbit each other with a very short period (often $<1$ d). Due to the relative brightness of the components, nearly all of the light in the system originates from the white dwarf. Because of the relative sizes of these objects and their light contributions, they exhibit transits in which a large amount of light of the white dwarf is blocked when the M-dwarf passes in front of the white dwarf. Additionally, due to their short orbital periods, these systems often spend very little time in transit, with transit times being of order minutes. Thus, these systems often have very few data points in transit, making them difficult to detect and their periods difficult to quantify, Here, we demonstrate that our nonparametric method, which inherently makes no assumption on the underlying morphology of variability, is able to robustly determine the frequency of variability of transit like signals in addition to eclipses and more sinusoidal-like variability. In Tab. \ref{tab:table_comparison}, we included three white dwarf plus M-dwarf binaries with published periods that were observed by ZTF \citep{Brown2023}. The FINKER results for ZTF J041016.82–083419.5 are shown in Fig. \ref{fig:kernen_regression_AMCV2}. We notice that while the shape of the regression does not completely match the transit, the methodology is robust identifying the correct period. We further note that a smaller bandwidth may be more appropriate when hunting for periodic phenomena that have extremely short durations in phase space.

\subsection{Radial Velocity time series}

In addition to photometric time series, binary stars and exoplanet systems experience periodic radial velocity shifts due to orbital motion. Except for dedicated cases, radial velocity time series are often sparsely sampled due to various scheduling and weather condition requirements. In addition to variability arising from orbital motion, several other phenomena can result in actual or apparent Doppler shifts at the stellar surface, including stellar pulsations \citep{Aerts2010}, rotation, and winds. As a result, there are often multiple sources of variability with different amplitudes in radial velocity measurements. In this work, we consider radial velocity time series for a series of known spectroscopic binaries that were obtained via spectroscopic observations with the HERMES Echelle spectrograph on the Mercator telescope in La Palma, Spain \citep{Raskin2011}. These data were obtained with various observing strategies and are spaced out over several months in some cases and several years in others. 

We find that this method is extremely powerful for searching sparsely sampled radial velocity time series for periodicities. In the cases listed in Table~\ref{tab:table_comparison}, the regression periodogram unambiguously identifies the orbital frequency despite the different cases having various amounts of data and signal to noise ratios for different measurements. Furthermore, as our method does not make assumptions on the morphology of the signal, it can identify periodic signals originating from both circular and eccentric orbits. Finally, while we do not attempt it here, this methodology has direct application to searching for periodic signals arising from exoplanets as well.

\begin{figure*}[h]
    \centering
    \includegraphics[width=.95\textwidth]{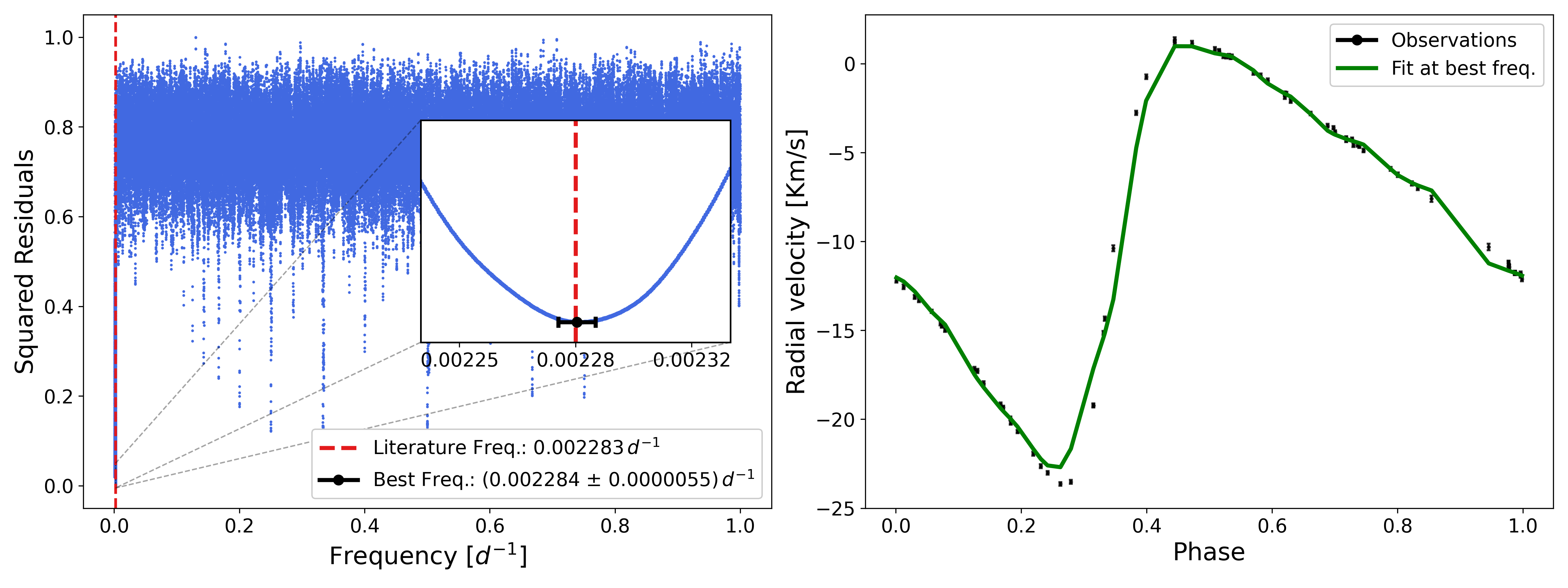}
    \caption{FINKER's frequency search for HD 114520. The figure on the left shows the frequency range searched, and the inset illustrates the smooth behaviour of the residuals around the minimum. The estimated best frequency and its uncertainty are shown with a black error bar. The figure on the right shows the light curve folded at the found frequency with the kernel regression fit (green line) overlaid, illustrating the algorithm's capability to model the periodic signal accurately.}
    \label{fig:folded_light_curve_specific_object}
\end{figure*}

\begin{figure*}[h]
    \centering
    \includegraphics[width=.95\textwidth]{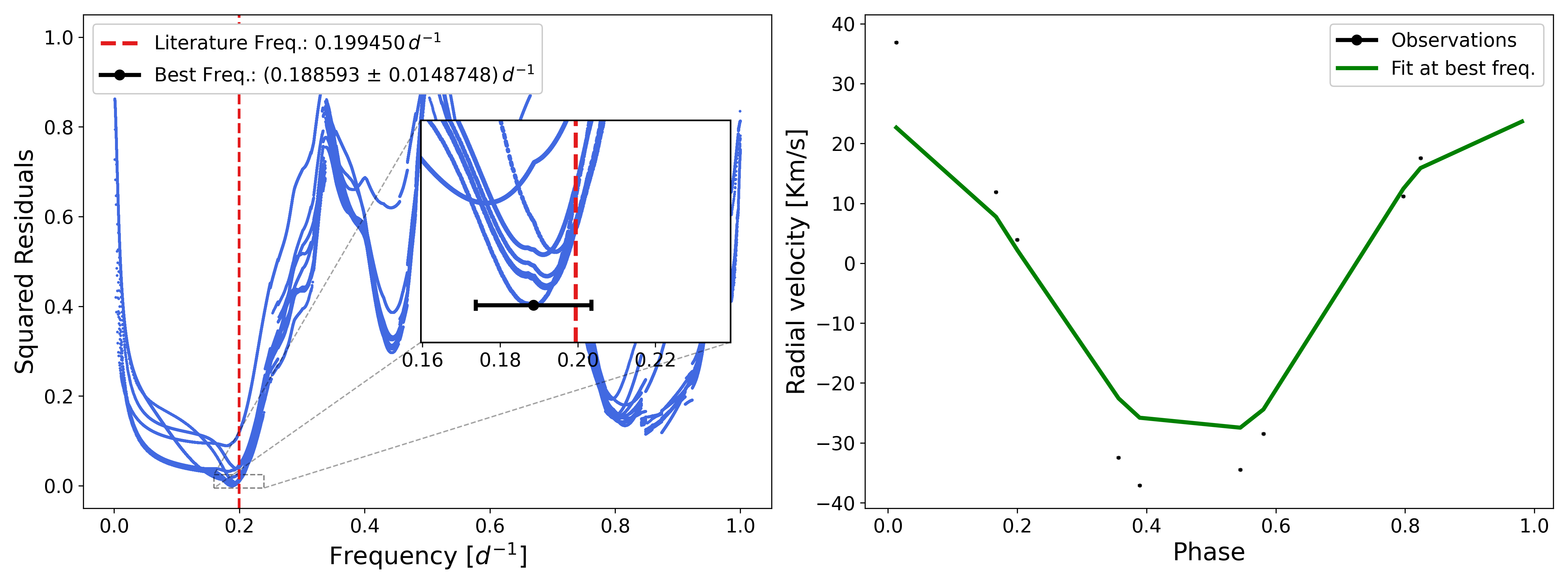}
    \caption{FINKER's frequency search for V772 Cas with an adaptive bandwidth. The figure on the left shows the frequency range searched and the wild behaviour of the residuals due to the low number of points. The inset shows that the best frequency agrees with the literature values; however, the estimated uncertainty is unreliable due to the really low number of observations. The figure on the right shows the light curve folded at the found frequency with the kernel regression fit (green line) overlaid.}
    \label{fig:folded_light_curve_specific_object}
\end{figure*}

\section{Conclusion and future outlooks}
\label{Sec: Conclusion}

This study has presented FINKER, a nonparametric periodogram for determining the frequencies of variability in astronomical time series, specifically sparsely sampled light curves and radial velocity measurement. At its core, this method performs kernel regression on a time series that has been folded over a proposed frequency and calculates the SSR of the data points with respect to the regression fit. A smaller SSR, in turn, means a more coherent structure and, therefore, corresponds to the likely period of variability. This method's main strength is its adaptability to localized data structures and inherent robustness to noise. These features enable this method to capture complex, non-sinusoidal periodic trends arising from numerous different physical mechanisms that faster parametric methods may miss. 

We evaluated and benchmarked our method by applying it to synthetic datasets, demonstrating a more accurate frequency estimation. Moreover, the robustness of kernel regression against the misidentification of harmonic frequencies has been highlighted as a key advantage over methods like conditional entropy. The empirical analysis of real-world datasets from the MeerLICHT telescope and other sources has further validated the algorithm's efficacy, showcasing its potential as a reliable tool for period determination in the field of astronomy.

Despite its strengths, the method is not without limitations. It is computationally more demanding than some traditional methods, which may restrict its use in processing extremely large datasets or in applications requiring real-time analysis. Additionally, while the method's susceptibility to half-frequency errors is mitigated by a simple check for significant doubled frequencies, this step requires additional computation and may not be foolproof in all cases. Specifically, this is the case for eclipsing binaries with similar eclipse depths. Future work will investigate whether a hybrid parametric-nonparametric method such as proposed by \citet{Saha2017} can mitigate this. Furthermore, there are clear failure cases when applying this method. These cases are limited to the situation where we do not achieve a quasi-uniform distribution of observations in the phase space for a give proposed frequency in the search. When this occurs, the regression can find several suitably similar configurations when comparing SSR values.  

Looking ahead, there are several promising directions for future research. First, we can investigate methods to increase the computational efficiency of this method. For example, a two-step approach using a faster method to first calculate a coarse frequency grid before using our kernel regression method to more finely sample the promising regions. Second, our method employed local constant regression, but we also tested local linear regression and found some improvement in complex features in phase space; however, as of now, its improvement does not outweigh the additional computational cost. We briefly discuss the use of a grid in phase space to speed up computations in Appendix~\ref{apdx:computation_efficiency}. 

\noindent
Future work could also concentrate on the use of measurement uncertainties directly in the kernel regression estimation, this is a complex problem still not completely solved in the statistics community. The use of errors-in-variable estimators and kernel deconvolution regression will be the first methods to explore in this regard \citep{Delaigle2006, Delaigle2014, DiMarzio2023}.
Furthermore, as the majority of stars exhibit more than a single periodic signal in photometry, expanding the algorithm's capabilities to automatically handle multi-periodic signals without manual intervention would be a valuable enhancement. 

\noindent
Finally, given the impending increase in data volume from upcoming multi-colour photometric missions such as BlackGEM \citep{Groot2019a,Groot2022} and Vera Rubin Observatory \citep{Rubin2019}, we will need new methods that are flexible and do not make assumptions on the underlying signals and that can efficiently search for periodicities in highly sparsely sampled data. This method promises to be an excellent application for use with such data.

\begin{acknowledgements}
CJ gratefully acknowledges support from the Netherlands Research School of Astronomy (NOVA) and from the Research Foundation Flanders (FWO) under grant agreement G0A2917N (BlackGEM). PJG is partly supported by SARChI Grant 111692 from the South African National Research Foundation. 
\end{acknowledgements}

\bibliographystyle{aa}
\bibliography{Bibliography}

\begin{appendix}

\section{Asymptotic properties}

The asymptotic properties of the nonparametric kernel regression models are critical for understanding the large-sample behaviour. These have been thoroughly studied in \citealp{Fan1993} and \citealp{Fan1994}. We summarise the key parameters: bias, variance, and mean squared error (MSE) to assess the asymptotic optimality of our methodology.

\paragraph{Bias}
For our local constant regression model, the bias is of $O(h^2)$, where $h$ is the bandwidth. In practice, the bias can be significantly reduced by choosing an optimal bandwidth through cross-validation methods.

\begin{equation}
\text{Bias}(\hat{f}_{\text{lc}}(x)) = O(h^2)
\end{equation}

\paragraph{Variance}
The variance of the estimator can be described as $O\left(\frac{1}{nh}\right)$ for both local constant and local linear models. 

\begin{equation}
\text{Variance}(\hat{f}_{\text{lc}}(x)) = O\left(\frac{1}{nh}\right)
\end{equation}

\paragraph{Mean Squared Error (MSE)}
The mean squared error is a function of both bias and variance and is given by:

\begin{equation}
\text{MSE}(\hat{f}_{\text{lc}}(x)) = O\left(h^2 + \frac{1}{nh}\right)
\end{equation}

\noindent
The optimal bandwidth minimizes this MSE, and cross-validation techniques are commonly employed to find this balance between bias and variance.

\paragraph{Asymptotic Normality}
As shown in \citealp{MartinsFilho2012O}, under mild regularity conditions, the local constant regression is asymptotically normal.
These asymptotic properties confirm that our nonparametric kernel regression models are statistically sound and efficient for large datasets.

\section{Computational efficiency of grid-based kernel regression}
\label{apdx:computation_efficiency}
Nonparametric kernel regression can be computationally intensive with large datasets. To improve efficiency, we tested a grid-based method. This approach evaluates the regression function at predetermined grid points rather than at every data point. We first create a grid over the data's range and compute the regression at these points. Then, we interpolate these values to estimate the regression function for the entire dataset.

\begin{figure}[h!]
\centering
\includegraphics[width=0.45\textwidth]{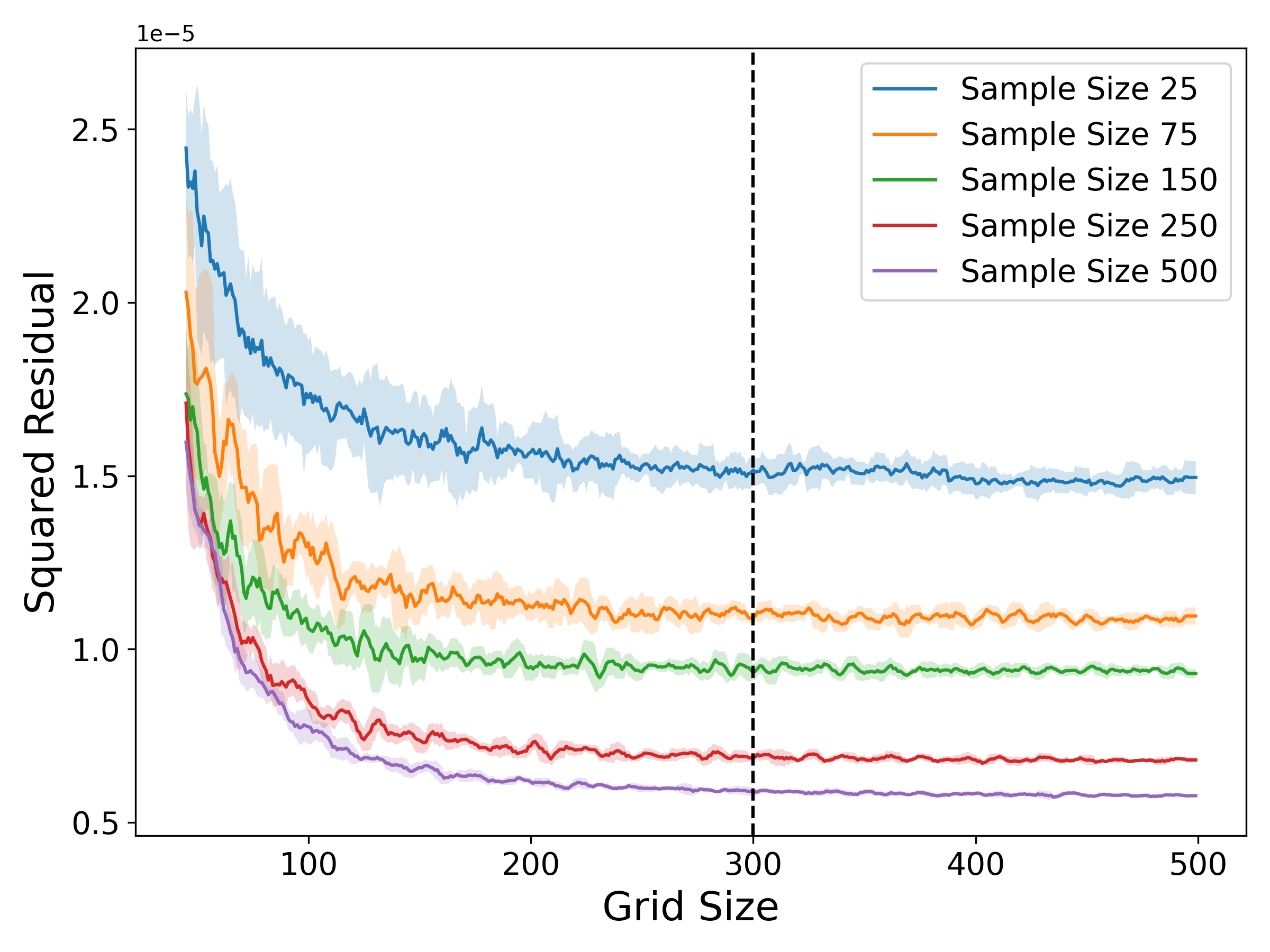}
\caption{Stability of squared residuals across various grid sizes, indicating a plateau at a grid size of approximately 300 points.}
\label{fig:squared_residuals_grid}
\end{figure}

We evaluated the computation times of kernel regression across various grid sizes using multiple synthetic light curves, both with and without the implementation of the grid method. 
The results showed that using a grid significantly reduces computation time, especially for large datasets. However, choosing the right grid size is important to ensure accuracy. We analyzed how the grid size affects the variability of residuals and found that a grid with about 300 points is sufficient for stable and consistent squared residuals, regardless of sample size, as shown in Figure \ref{fig:squared_residuals_grid}.

\begin{figure}[h!]
\centering
\includegraphics[width=0.45\textwidth]{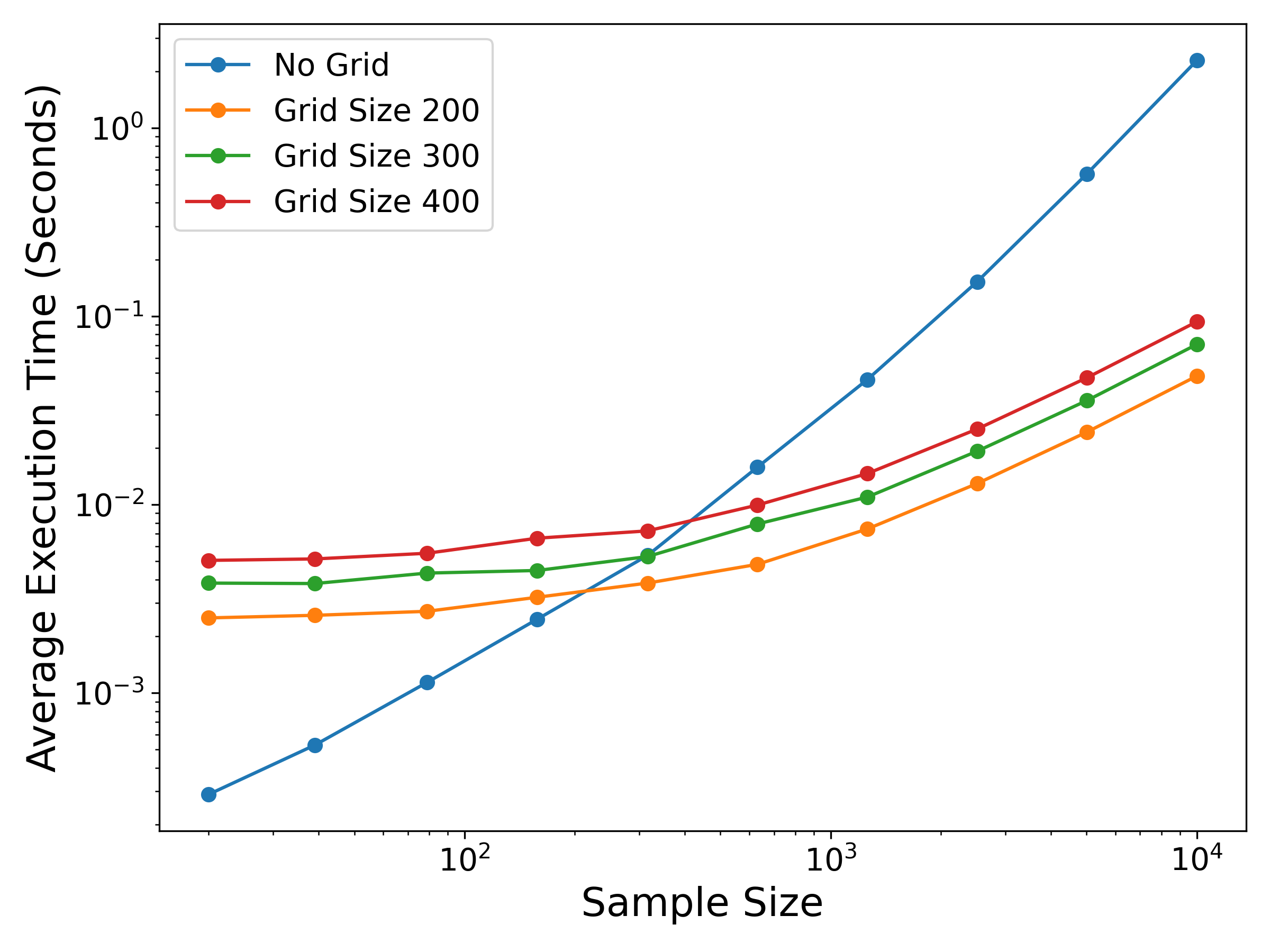}
\caption{Execution times comparing direct evaluation of the kernel regression on the points and on grids of different sizes, showing efficiency gains with grid sizes beyond 300 points.}
\label{fig:execution_time_grid}
\end{figure}

\noindent
Figure \ref{fig:execution_time_grid} presents the average computation times without grid and with grids of different sizes. For smaller sample sizes, direct evaluation on the observations is quicker. But for larger samples, using a grid speeds up the process without affecting the regression's accuracy.

%In summary, the grid method is a practical way to speed up kernel regression for large datasets without losing accuracy in the results.

\noindent
All tests were run on an Alienware Area 51M, Intel Core i9-9900K, 32GB DDR4/2400, Nvidia GeForce RTX 2080.

\section{Controversial cases}

In this appendix, we delve into specific instances where FINKER encountered challenges or yielded results that diverged from established literature. 

\paragraph{Case Study: KIC 5217733}

\begin{figure*}[h]
    \centering
    \includegraphics[width=.95\textwidth]{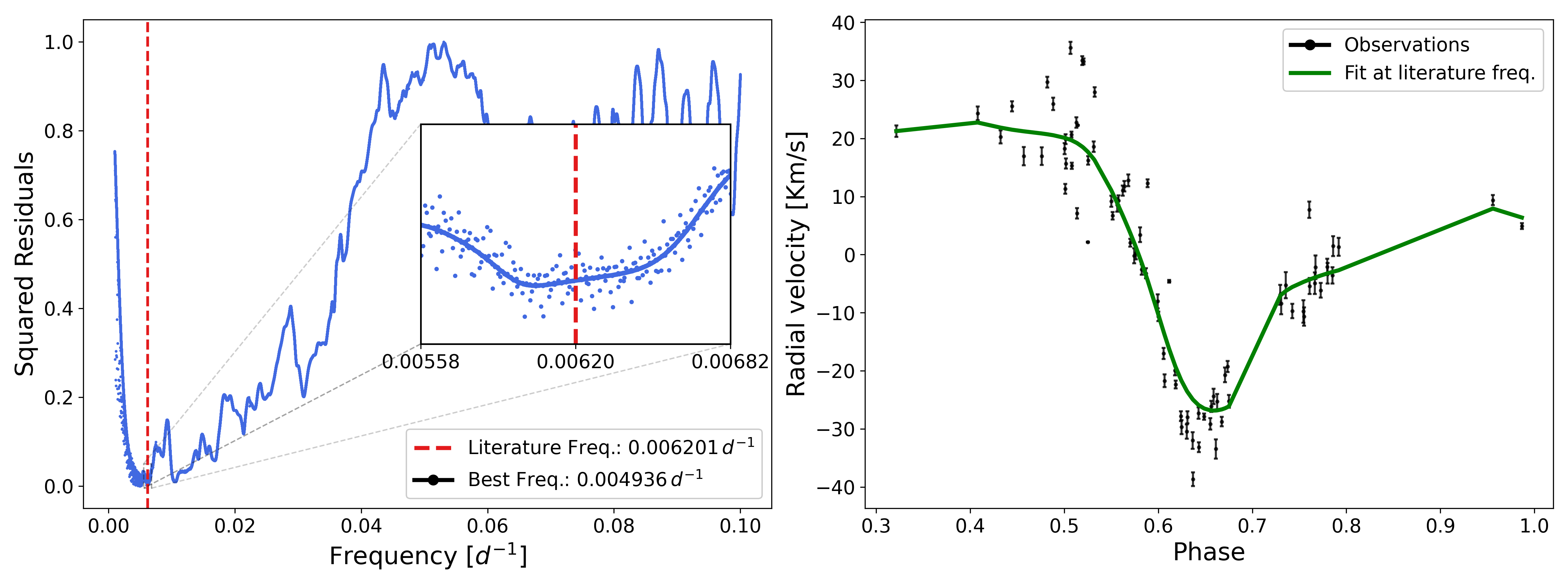}
    \caption{FINKER's frequency search for KIC 5217733  \citep{Kirk2016}. The figure on the left shows the frequency range searched, and the inset illustrates the failed recovery of the literature frequency. The figure on the right shows the light curve folded at the literature frequency ($0.006201 d^{-1}$) with the kernel regression fit (green line) overlaid.}
    \label{fig:Example_KIC}
\end{figure*}

For KIC 5217733, FINKER's frequency identification process did not align with the frequencies reported in literature \citep{Kirk2016}. As shown in Figure \ref{fig:Example_KIC}, the algorithm's search spanned the expected frequency range but failed to accurately recover the known frequency. This discrepancy is possibly attributed to the underestimation of uncertainties and the presence of numerous g-mode pulsations, which introduced additional variations in the line profile.

\paragraph{Case Study: MLT J162036.64--614110.5}

\begin{figure*}[h]
    \centering
    \includegraphics[width=.95\textwidth]{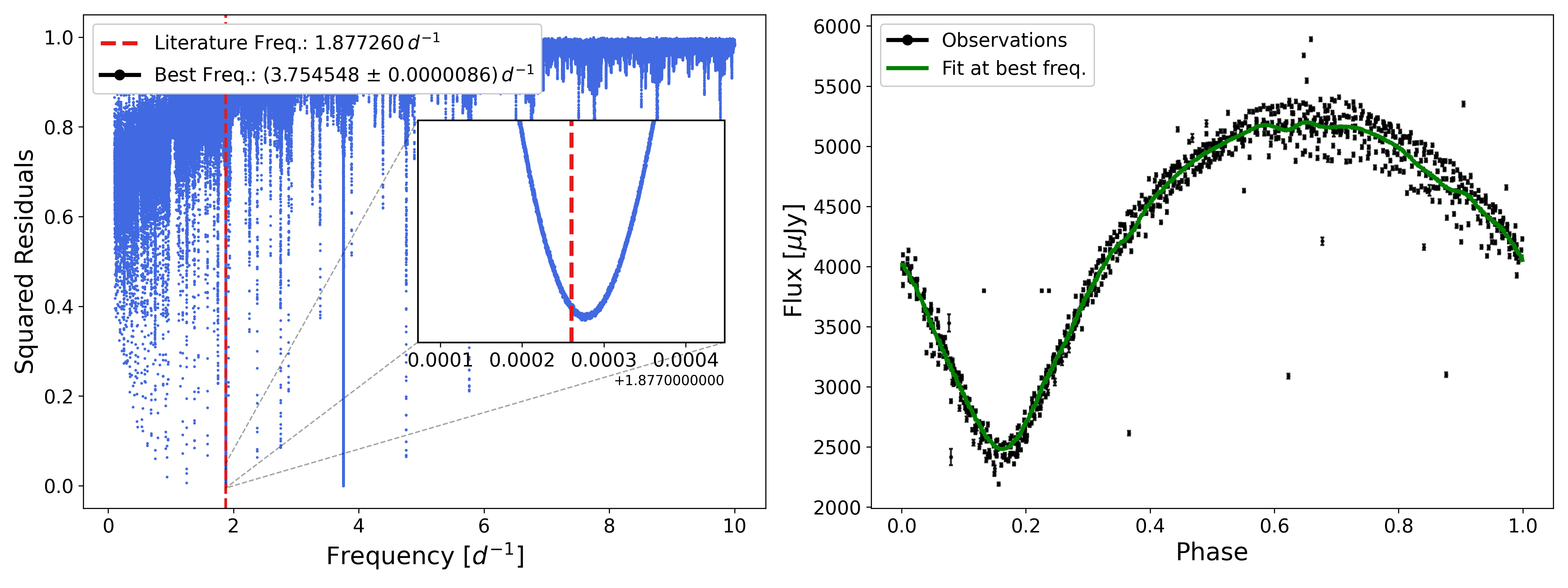}
    \caption{FINKER's frequency search for MLT J162036.64--614110.5. The figure on the left shows the frequency range searched and the thorough evaluation around the best frequency and its subharmonic. In disagreement with the results in the literature, FINKER did not find a significant difference in the eclipses. The figure on the right shows the light curve folded at the found frequency with the kernel regression fit (green line) overlaid.}
    \label{fig:kernel_regression_EB_2}
\end{figure*}

Another intriguing case is MLT J162036.64--614110.5, where FINKER's results diverged from established literature findings. As depicted in Figure \ref{fig:kernel_regression_EB_2}, the algorithm executed a comprehensive search, including analyses of both the main frequency it found and its subharmonic. Interestingly, FINKER did not observe any significant difference in the eclipses, in contrast to literature reports. The light curve, when folded at the frequency identified by FINKER, and overlaid with our kernel regression fit, suggests a different interpretation of the data compared to traditional analyses. This case underscores the potential for FINKER's framework to provide alternative insights into celestial phenomena, although it also highlights the need for cautious interpretation, especially in instances where results significantly deviate from established knowledge.

\end{appendix}
\end{document}